\documentclass[table,sigplan,screen]{acmart}
\usepackage[T1]{tipa}
\usepackage[tt=false, type1=true]{libertine}
\usepackage[libertine]{newtxmath}

\usepackage{tikz,pgfplots}
\usetikzlibrary{arrows.meta}

\pgfplotsset{every axis/.append style={font=\sffamily\scriptsize}}
\usepackage{float}
\usepackage{tipa}
\usepackage{graphicx}
\usepackage{balance}
\usepackage{stfloats}
\usepackage[caption=false,font=footnotesize,labelfont=sf,textfont=sf]{subfig}
\usepackage{booktabs}
\usepackage{makecell}
\usepackage{mdframed}
\usepackage{listings}
\usepackage{bm}
\usepackage{enumitem}
\usepackage{multirow}
\usepackage{nicefrac}
\usepackage{amsmath}
\usepackage{amssymb}
\usepackage{booktabs}
\usepackage{todonotes}
\usepackage{arydshln}
\usepackage{booktabs,xcolor,siunitx}

\usepackage{xcolor}
\usepackage{tcolorbox}
\definecolor{lightgrey}{HTML}{D3D3D3}

\def\BibTeX{{\rm B\kern-.05em{\sc i\kern-.025em b}\kern-.08emT\kern-.1667em\lower.7ex\hbox{E}\kern-.125emX}}

\copyrightyear{2020}
\acmYear{2020}
\setcopyright{acmcopyright}
\acmConference[CGO '20]{Proceedings of the 18th ACM/IEEE International Symposium on Code Generation and Optimization}{February 22--26, 2020}{San Diego, CA, USA}
\acmBooktitle{Proceedings of the 18th ACM/IEEE International Symposium on Code Generation and Optimization (CGO '20), February 22--26, 2020, San Diego, CA, USA}
\acmPrice{15.00}
\acmDOI{10.1145/3368826.3377904}
\acmISBN{978-1-4503-7047-9/20/02}

\emergencystretch=\maxdimen
\begin{document}

\title[AN5D: Automated Stencil Framework for High-Degree Temporal Blocking...]{AN5D: Automated Stencil Framework for High-Degree Temporal Blocking on GPUs}

\settopmatter{authorsperrow=3}

\author{Kazuaki Matsumura}
\affiliation{\hspace*{-7pt}\mbox{Barcelona Supercomputing Center \country{Spain}}}
\email{kazuaki.matsumura@bsc.es}
\authornotemark[1]

\author{Hamid Reza Zohouri}
\affiliation{Edgecortix Inc. \country{Japan}}
\email{hamid@edgecortix.com}
\authornotemark[1]

\author{Mohamed Wahib}
\affiliation{AIST \country{Japan}}
\email{mohamed.attia@aist.go.jp}

\author{Toshio Endo}
\affiliation{Tokyo Institute of Technology \country{Japan}}
\email{endo@is.titech.ac.jp}

\author{Satoshi Matsuoka}
\affiliation{RIKEN CCS \country{Japan}}
\email{matsu@acm.org}

\renewcommand{\shortauthors}{K. Matsumura, H. R. Zohouri, M. Wahib, T. Endo, S. Matsuoka}

\setlength{\textfloatsep}{10pt}
\setlength{\dbltextfloatsep}{10pt}
\setlength{\abovedisplayskip}{0pt}
\setlength{\belowdisplayskip}{6pt}
\setlength{\abovedisplayshortskip}{0pt}
\setlength{\belowdisplayshortskip}{6pt}
\begin{abstract}
Stencil computation is one of the most widely-used compute patterns in high performance computing applications. Spatial and temporal blocking have been proposed to overcome the memory-bound nature of this type of computation by moving memory pressure from external memory to on-chip memory on GPUs. However, correctly implementing those optimizations while considering the complexity of
the architecture and memory hierarchy of GPUs to achieve high performance is difficult.
We propose AN5D, an automated stencil framework which is capable of automatically transforming
and optimizing stencil patterns in a given C source code, and generating corresponding CUDA code.
Parameter tuning in our framework is guided by our performance model. Our novel optimization strategy
reduces shared memory and register pressure in comparison to existing implementations, allowing performance
scaling up to a temporal blocking degree of 10. We achieve the highest performance reported so far for all evaluated stencil benchmarks on the state-of-the-art Tesla V100 GPU.
\end{abstract}



\maketitle

\section{Introduction}\label{sec:introduction}

Stencil computation is one of the most frequently used computation patterns
in High Performance Computing (HPC) that is often highly iterative~\cite{shimokawabe201080, shimokawabe2011peta, Rossinelli:2013:PSC:2503210.2504565, 7967144}.
Despite non-stop advancements in both hardware and compiler technologies,
optimizing this computation pattern on modern hardware remains a challenge
due to its memory-bound nature. Currently, GPUs are the most popular accelerator
in supercomputers and are employed in half of the top 10
machines in the TOP500 June 2019 list~\cite{top500}.
Although
the computational performance of GPUs has been increasing rapidly,
the gap between their computational performance and
memory bandwidth prevents full utilization of their computational
performance for HPC applications that rely on stencil computation.
Temporal blocking~\cite{Krishnamoorthy:2007:EAP:1250734.1250761, Meng:2009:PMA:1542275.1542313, Holewinski:2012:HCG:2304576.2304619, maruyama2014optimizing, Rawat:2015:SMD:2830018.2830025, grosser2013promises, prajapati2017simple, 3.5d, stencilgen1, stencilgen2}
is a well-known technique proposed to relieve memory pressure in stencil computation
by combining multiple consecutive iterations of the time loop
and avoiding the memory transactions required between them.
However, correct and efficient implementation of this technique is challenging
since it requires careful management of limited register and shared memory resources on GPUs.
As a result, most existing work that implement temporal blocking is limited
to low
degrees of temporal blocking. Moreover, such work is generally implemented
manually on a limited set of benchmarks~\cite{maruyama2014optimizing,3.5d},
or through frameworks that are not available for public use~\cite{Rawat:2015:SMD:2830018.2830025, stencilgen1, stencilgen2}.
{\let\thefootnote\relax\footnote{{*This work was performed while those authors were at Tokyo Institute of Technology and K. Matsumura was a research assistant at RWBC-OIL.}}}
In this work, we present our open-source
high-performance stencil code generation framework called {\bf AN5D} \textipa{["{\ae}ndi]} ({\bf A}uto \mbox{{\bf N}.{\bf 5}{\bf D}}).
Our framework accepts unoptimized stencil patterns in C language,
implements spatial and temporal blocking alongside with multiple low-level optimizations,
and generates associated CUDA host and kernel code.
Our contributions are as follows:
\begin{itemize}[topsep=0.5em, leftmargin=0.1in,rightmargin=0.05in, itemsep=2pt]
\item We create an automated stencil framework
to automatically implement spatial and temporal blocking
from a C-based stencil description and automatically generate
associated CUDA code.
\item Through substantial engineering effort,
we implement multiple low-level optimization techniques
such as associative stencil optimization, shared memory double-buffering
to reduce shared memory usage, and data movement reduction to optimize register usage,
\linebreak
all in a cohesive fashion. This allows us to scale performance with temporal
blocking degrees up to 10 on GPUs, which had never been achieved before.
\item We perform a comprehensive comparison with state-of-the-art
implementations of stencil computation on GPUs (Hybrid Tiling and deep temporal tiling)
and show that our framework achieves the highest performance reported so far
for a wide range of stencil shapes and orders on the latest NVIDIA GPUs.
\item We make our framework available to public to relieve the community from the
burden of having to implement all those optimizations manually.
\end{itemize}
 \vspace*{-0.3cm}
\section{Background}

\subsection{Stencil Computation}
In stencil computation, grid cells from a multi-dimensional
input are iteratively updated based on a specific computation pattern;
this pattern is defined by the stencil \textit{shape}. The most widely used
stencil shapes are \textit{star} and \textit{box}. Star stencils
involve accessing neighbors that differ from the current cell
only in the direction of one dimension at a time (i.e. no diagonal accesses),
while box stencils form full squares (for 2D) or cubes (for 3D).
The computation pattern involves neighboring accesses along
each dimension up to a certain distance called the stencil \textit{radius}.
We call a stencil with radius of $rad$ a ``$rad^\mathrm{th}$-order stencil''\footnote{Some
scientific publications call such stencil a ``$(2 \times rad)^\mathrm{th}$-order stencil''}.
The calculation generally involves a sum of products over the values of the
current cell and its neighbors and a set of coefficients that might or might
not be shared among these cells. Therefore, each calculation depends on the
calculation of neighbors in the previous iteration (also called \textit{time-step}).
\subsection{Spatial Blocking}
Since there is no data dependency within the same time-step for explicit solvers,
the spatial order of stencil updates within one time-step can be arbitrary.
Spatial blocking~\cite{Wolfe:1989:MIS:76263.76337, irigoin1988supernode}
splits the input grid into multiple blocks (or sub-planes)
to accelerate execution by increasing the locality.
On GPUs, rather than purely relying on the cache hierarchy,
on-chip resources can be used to explicitly implement spatial blocking.
After one block of data is moved from global memory to on-chip memory,
the remaining loads in the time-step from the same block are done with no
global memory access. However, loading boarder neighbor cells which belong
to adjacent blocks will result in redundant global memory accesses.

\subsection{Temporal Blocking}
Even though stencil computation has data dependency across time-steps,
the dependency range of one cell is limited by the product of the stencil radius $({rad})$
and the number of time-steps passed since the cell's last update $(b_{\mathrm{T}})$.
Temporal blocking exploits the hidden temporal locality by combining multiple consecutive time-steps
and avoiding global memory accesses in between. The dependency along the time dimension is resolved
by redundantly loading and computing cells from adjacent blocks, the amount of which will increase
with ${rad}$ and $b_{\mathrm{T}}$.

\begin{figure}[b]
  \centering
  \includegraphics[trim=0 0.25cm 0 0.25cm,clip,width=\linewidth]{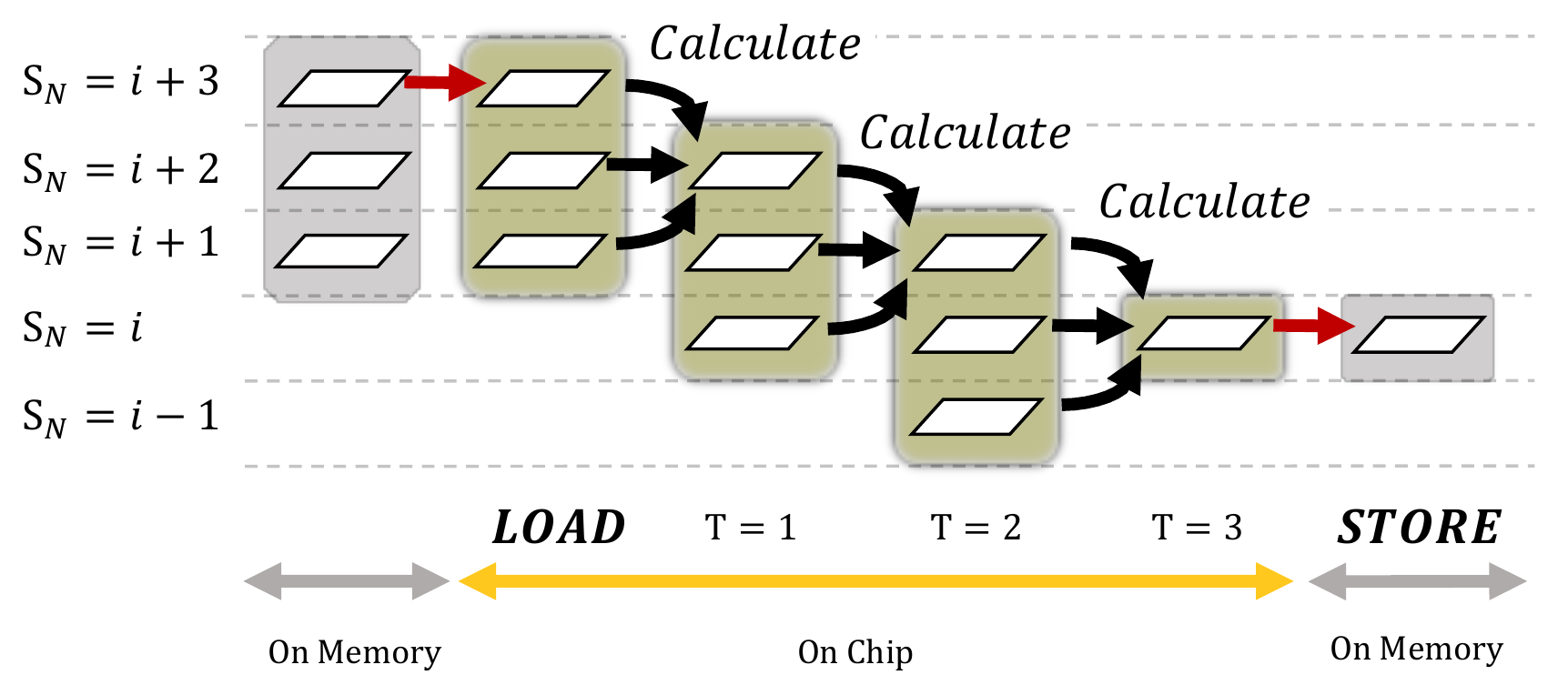}
  \vspace{-0.6cm}
  \caption{\small Computational flow of N.5D blocking with temporal blocking size of $3$ and stencil radius of $1$}
  \label{fig:tier}
\end{figure}

Overlapped tiling~\cite{Krishnamoorthy:2007:EAP:1250734.1250761, Meng:2009:PMA:1542275.1542313, Holewinski:2012:HCG:2304576.2304619, Rawat:2015:SMD:2830018.2830025}
is a form of temporal blocking which involves \textit{overlapping} the spatial blocks
by $2 \times b_{\mathrm{T}} \times {rad}$ rows and columns (called \textit{halo} regions)
and redundantly loading and computing necessary cells that would fall inside surrounding blocks to process $b_{\mathrm{T}}$
time-steps with only one round of global memory loads and stores per block. Overlapped tiling is also applicable
\linebreak
over spatial blocking methods other than domain decomposition (i.e. blocking \textit{all} the input dimensions).
3.5D blocking, which implements 1D overlapped tiling (2 combined time-steps) on top of 2.5D spatial blocking for
3D stencils, was introduced in ~\cite{3.5d}. 2.5D spatial blocking involves blocking two dimensions and streaming the
computation over the third one. Similarly, 1.5D spatial blocking can be used for 2D stencils. 
This blocking technique has been widely employed on different devices ~\cite{maruyama2014optimizing,Rawat:2015:SMD:2830018.2830025,zohouri2018combined,zohouri2018high},
with not just two, but also more combined time-steps.
In 2.5D spatial blocking, the 2D tiles are streamed over one dimension and data of each tile is effectively reused for updating the next tiles.
To minimize the use of on-chip memory, the computational flow forms tiers along time-steps as shown in Fig.~\ref{fig:tier};
here, each tile is evicted from on-chip memory as soon as no computation depends on its data anymore,
and is replaced by a new tile. 3.5D blocking can be extended to support any arbitrary number of combined time-steps;
we call this \textit{N.5D blocking}.

In contrast to overlapped tiling, non-redundant temporal blocking techniques~\cite{Kamil:2006:IEO:1178597.1178605, grosser2013split, muranushi2015optimal, bondhugula2017diamond}
realize temporal blocking without introducing redundant computation.
To resolve the temporal dependency, such tiling methods form geometric (e.g. trapezoidal, wavefront) blocks
along the time dimension. However, the dependency between neighboring blocks restricts parallelism in those methods.
 \vspace*{-0.3cm}
\section{Related Work}
\label{related}
In~\cite{stencilgen1,stencilgen2}, Rawat et al. present a DSL-based stencil framework named
STENCILGEN that implements N.5D blocking with several optimizations including DAG fusion.
In this framework, shared memory is used for accessing nearby cells from each thread within a block, and one shared memory buffer is required per combined time-step.
For diagonal-access free (a.k.a star) stencils,
registers are used to keep the previous time-step's results of both the upper and the lower sub-planes.
For other stencil types, if the stencil is \textit{associative},
the computation of each cell is carried out in multiple steps, with each step only
performing the parts of the computation that access one sub-plane (partial summation).
This way, it is not required to keep all sub-planes that need to be accessed for computation
of a cell in shared memory at the same time and instead, sub-planes are computed one
by one and the result of the partial sum is stored in a register. Each partial sum is then used for computing
the next sub-plane to complete the summation. This technique was also used in~\cite{6569833} to improve
global memory access alignment and efficiency. Moreover, STENCILGEN supports the division of the streaming
dimension for increasing thread-block-level parallelism, at the cost of additional redundant computations
along the streaming dimension. Main advantages of our framework over STENCILGEN are:  1) Our framework uses standard C code as input while STENCILGEN is DSL-based.
2) Our framework is generic and publicly available (some benchmarks optimized by STENCILGEN are publicly available but not the framework itself).
3) Our framework reduces both shared memory and register usage by reducing register movement and employing
shared memory double-buffering instead of multi-buffering (Section~\ref{execmodel} and Table~\ref{tab:comp_to_stencilgen}), 4) And most importantly,
our novel optimization strategy allows performance scaling up to two-digit degrees of temporal blocking,
while STENCILGEN's performance scaling is limited to a degree of 4~\cite{stencilgen1,stencilgen2}.
In~\cite{rawatoptimizing}, Rawat et al. present another DSL-based stencil framework called ARTEMIS
which supports flexible resource allocation on GPUs (global memory or share memory + register)
for each input/output grid especially for high-order multi-array and multi-statement stencils.
In~\cite{Rawat:2018:ROS:3178487.3178500}, the same authors present a statement reordering framework
for complex high-order stencils that avoids using shared memory and minimizes register pressure
by using a register scheduling algorithm to re-arrange the sequence of memory access/compute statements.
In contrast, AN5D focuses on low-level optimizations to enable high-degree
temporal blocking that could complement such optimizations.

\begin{figure}[t]
  \centering
  \includegraphics[width=0.93\linewidth]{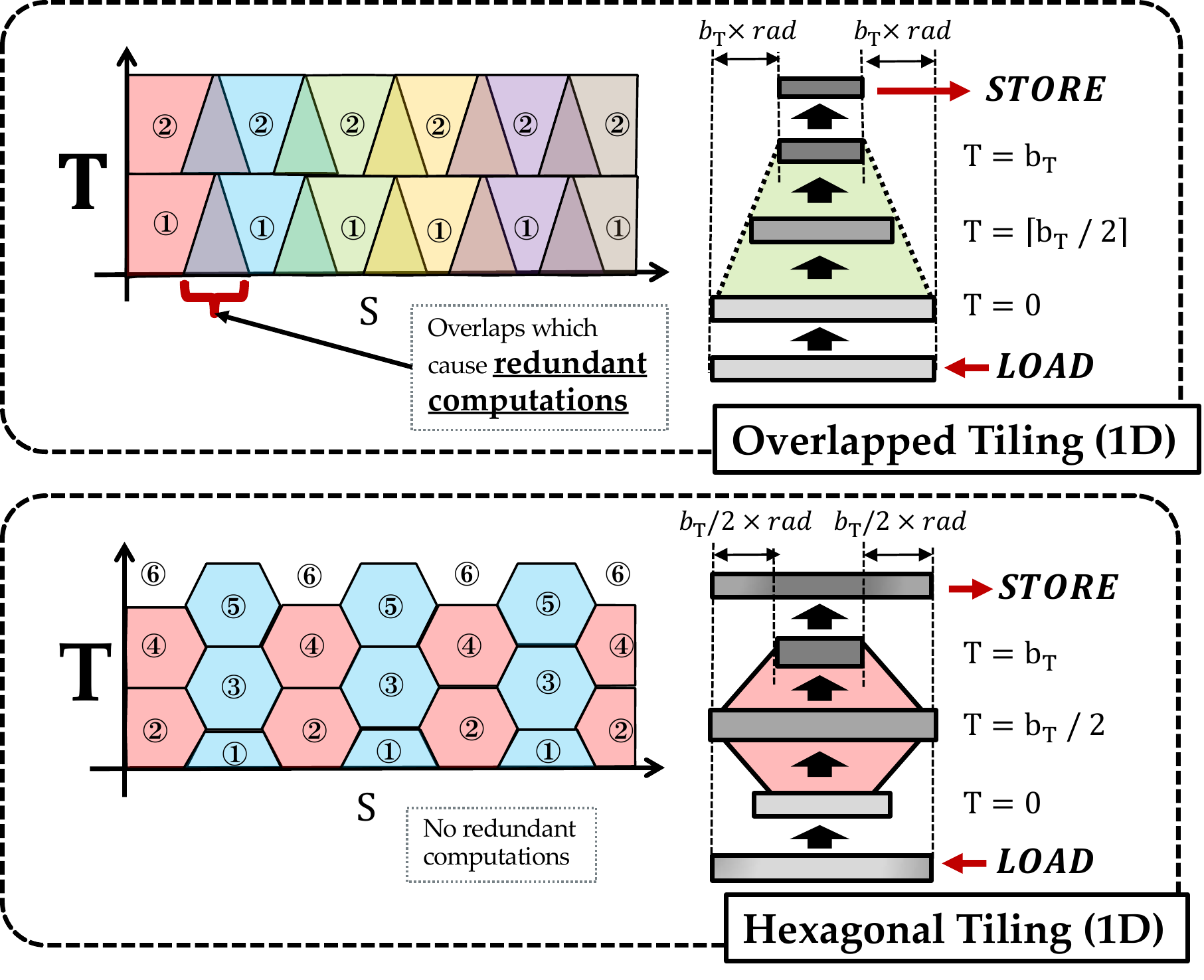}
  \vspace{-0.3cm}
  \caption{\small Computational flow of 1D overlapped and hexagonal tiling. Red arrows indicate global memory I/O and black arrows depict the update flow that involves on-chip memory accesses.}
  \label{fig:1d}
\end{figure}

{
\renewcommand{\arraystretch}{1.15}
\setlength{\tabcolsep}{2pt}
\newcolumntype{L}{ >{\raggedright\let\newline\\\arraybackslash} m }
\newcolumntype{M}{ >{\centering\let\newline\\\arraybackslash} m }
\begin{table}[b]
  \centering
  \caption{\small Comparison to STENCILGEN}
  \label{tab:comp_to_stencilgen}
  \vspace{-0.35cm}
  \footnotesize\begin{tabular}{L{2.5cm} M{3cm} M{2.2cm}}
    \Xhline{2\arrayrulewidth}
     & STENCILGEN~\cite{stencilgen2} & AN5D
    \\
    \Xhline{2\arrayrulewidth}
    {Register Allocation}  & Shifting & Fixed\\
    {Shared Memory Use} & For Streaming & For Calculation\\
    \hline
    \multicolumn{2}{l}{{\bf Shared Memory Footprint Per Block}:}  & \\
    \textit{Diagonal-Access Free}  
                   & $n_{\mathrm{thr}} \times b_{\mathrm{T}} \times  n_{\mathrm{word}} $ &
                                                $ 2 \times  n_{\mathrm{thr}} \times  n_{\mathrm{word}}$ \\
    \textit{Associative Stencil}  & $n_{\mathrm{thr}} \times b_{\mathrm{T}} \times  n_{\mathrm{word}} $ &  $ 2 \times  n_{\mathrm{thr}} \times  n_{\mathrm{word}}$ \\
    \multirow{2}{*}{\textit{Otherwise}} & $n_{\mathrm{thr}} \times b_{\mathrm{T}} \times$ & $2 \times  n_{\mathrm{thr}} \times$  \\
    & $ (1 + 2 \times {rad}) \times n_{\mathrm{word}} $ & $ (1 + 2 \times {rad}) \times n_{\mathrm{word}} $ \\
    \hline
    \multicolumn{2}{l}{{\bf Shared Memory Store Per Cell}:}  & \\
    \textit{Diagonal-Access Free} & $1$ & $ 1 $ \\
    \textit{Associative Stencil} & $1$ & $ 1 $ \\
    \textit{Otherwise} & $ 1 + 2 \times {rad}$ & $ 1 + 2 \times {rad}$ \\
\Xhline{2\arrayrulewidth}
  \end{tabular}
~\\
~\\
  {
(~~~$n_{\mathrm{word}}$~: The number of words for each cell value~ )%
  }
  \vspace*{-1em}
\end{table}
}

Hybrid tiling~\cite{grosser2013promises, hybrid,grosser2014relation,prajapati2017simple} is a prominent method
for non-redundant temporal blocking which combines \textit{hexagonal tiling} and classical wavefront tiling.
The computational flow of hexagonal tiling is shown in Fig.~\ref{fig:1d}.
This schedule (~\textcircled{\scriptsize 1}~$\rightarrow$~\textcircled{\scriptsize 2}~$\rightarrow$~\textcircled{\scriptsize 3}~$\rightarrow$~...~)
allows discrete blocks to be executed in parallel while the hexagonal shape allows resolving the temporal dependency without redundant computation.
Apart from the time dimension, hybrid tiling employs hexagonal tiling only for one of the spatial dimensions,
while remaining dimensions are blocked in a wavefront manner.
The main shortcoming of this technique is that it blocks all spatial dimensions (no streaming) and hence,
is limited to smaller block sizes compared to N.5D blocking for the same amount of on-chip memory.
This leads to higher ratio of redundant-to-valid memory accesses and lower scaling with temporal blocking
compared to N.5D blocking (mathematical proof provided in~\cite{3.5d}) on the same hardware.

Overtile~\cite{Holewinski:2012:HCG:2304576.2304619}, Forma~\cite{Ravishankar:2015:FDI:2716282.2716290} and SDSLc~\cite{Rawat:2015:SMD:2830018.2830025}
are other frameworks that can accelerate stencil computation by overlapped tiling on GPUs; however, none of them employs dimension streaming.
Among non-redundant temporal blocking techniques, trapezoidal tiling~\cite{grosser2013split},
diamond tiling~\cite{7582549}, and combined diamond and hexagonal tiling~\cite{grosser2014relation}
have been proposed.
Pochoir~\cite{Tang:2011:PSC:1989493.1989508} and YASK~\cite{7836083} are also stencil frameworks
that conduct temporal blocking for CPUs and Xeon Phi, respectively.
LIFT~\cite{Hagedorn:2018:HPS:3179541.3168824} is a functional data-parallel programming language that allows expressing
stencil loops as a set of reusable parallel primitives and optimizing them.
Recently, multiple implementations of N.5D blocking on FPGAs have also been proposed
with FPGA-specific optimizations~\cite{zohouri2018combined, zohouri2018high, torstenfpgastencil, congstencilfpga}.
FPGAs tend to achieve better scaling with temporal blocking compared to GPUs due to
higher flexibility of employing their on-chip memory which allows larger spatial block sizes compared to GPUs.
However, their final performance still falls short of that of modern GPUs due to large gap in peak compute and memory performance.

 \vspace*{-0.1cm}
\section{AN5D Framework}

\subsection{Execution Model}
\label{execmodel}

We denote the temporal blocking size (number of combined iterations/time-steps) as $b_{\mathrm{T}}$, and
the spatial blocking (sub-plane) size as $b_{\mathrm{S}_i}$ along each spatial dimension ($\mathrm{S}_i;\ 1 \leq i < {N}$)
which excludes the streaming dimension. Since each thread only processes one cell per block,
the thread-block size ($n_{\mathrm{thr}}$) used by our framework will be $n_{\mathrm{thr}} = \prod_{i=1}^{{N}-1} b_{\mathrm{S}_i}$.

The number of threads per dimension that store updated cells to global memory is represented by $b_{{\mathrm{S}_i}} - 2 \times b_{\mathrm{T}} \times {rad}$;
the difference from $n_{\mathrm{thr}}$ indicates the overlapped (halo) area for each block. We call the non-overlapped region the \textit{compute region};
these regions cover the entire input grid. The total number of thread-blocks ($n_{\mathrm{tb}}$) required for computation is given by 
$n_{\mathrm{tb}} = \prod_{i=1}^{{N}-1} \left\lceil \frac{{I}_{\mathrm{S}_i}}{b_{{\mathrm{S}_i}} - 2 \times b_{\mathrm{T}} \times {rad}} \right\rceil$.
Here, we denote the number of iterations for the spatial dimensions (i.e. input grid size) and the time dimension as ${I}_{\mathrm{S}_i}$ and ${I}_{\mathrm{T}}$, respectively.

A sub-plane traverses the streaming dimension from bottom to top,
while being accompanied by $b_{\mathrm{T}}$ computational streams, all of which are processed by the same thread-block.
Each computational stream undertakes computation of one of the combined time-steps and follows the behavior of the previous time-step's stream.
Computing each sub-plane depends on $1 + 2 \times {rad}$ sub-planes from the previous time-step.
One of these sub-planes is in the same ${N}$-th spatial position as the current one and rest are the upper and the lower sub-planes.
Thus, we store $1 + 2 \times {rad}$ cells in registers for each time-step and for each thread.
The streaming latency between time-steps can be seen in Fig.~\ref{fig:tier}.
After the first stream ($\mathrm{T}=0$) reads one sub-plane from global memory,
each next time-step updates a successive sub-plane with the distance of ${rad}$ sub-planes.
The size of the region with valid computation, defined as $\prod_{i=1}^{{N-1}} \left( b_{\mathrm{S}_i} - 2 \times T \times {rad} \right)$,
gets smaller in accordance with the increased time-step ($0 < T \leqslant b_{\mathrm{T}}$),
and the last computational stream ($\mathrm{T}=b_{\mathrm{T}}$) stores the results to global memory only for the non-overlapped area (compute region).

To avoid unnecessary branching and thread-diverges for threads falling inside the halo regions,
AN5D overwrites data in these regions with their original values.
Furthermore, at the start of the streaming, we use the register variables that
are prepared for storing the results of the $\mathrm{T}=b_{\mathrm{T}}-1$ computational
stream to maintain the first ${rad}$ sub-planes on the non-computational space;
these sub-planes include the boundary conditions (neighbors for boundary cells).
This eliminates the need to reload the same sub-planes from global memory on later time-steps.
For the end of the streaming, we instead use the register variables of $\mathrm{T}=0$
to hold these constant boundary sub-planes.

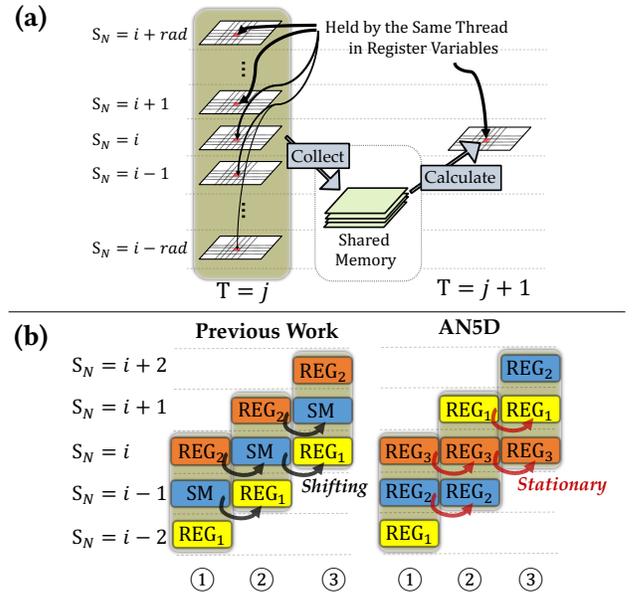
\begin{figure}[b]
  \vspace*{-0.12cm}
  \centering
  \subfloat{\includegraphics[width=0.34\textwidth]{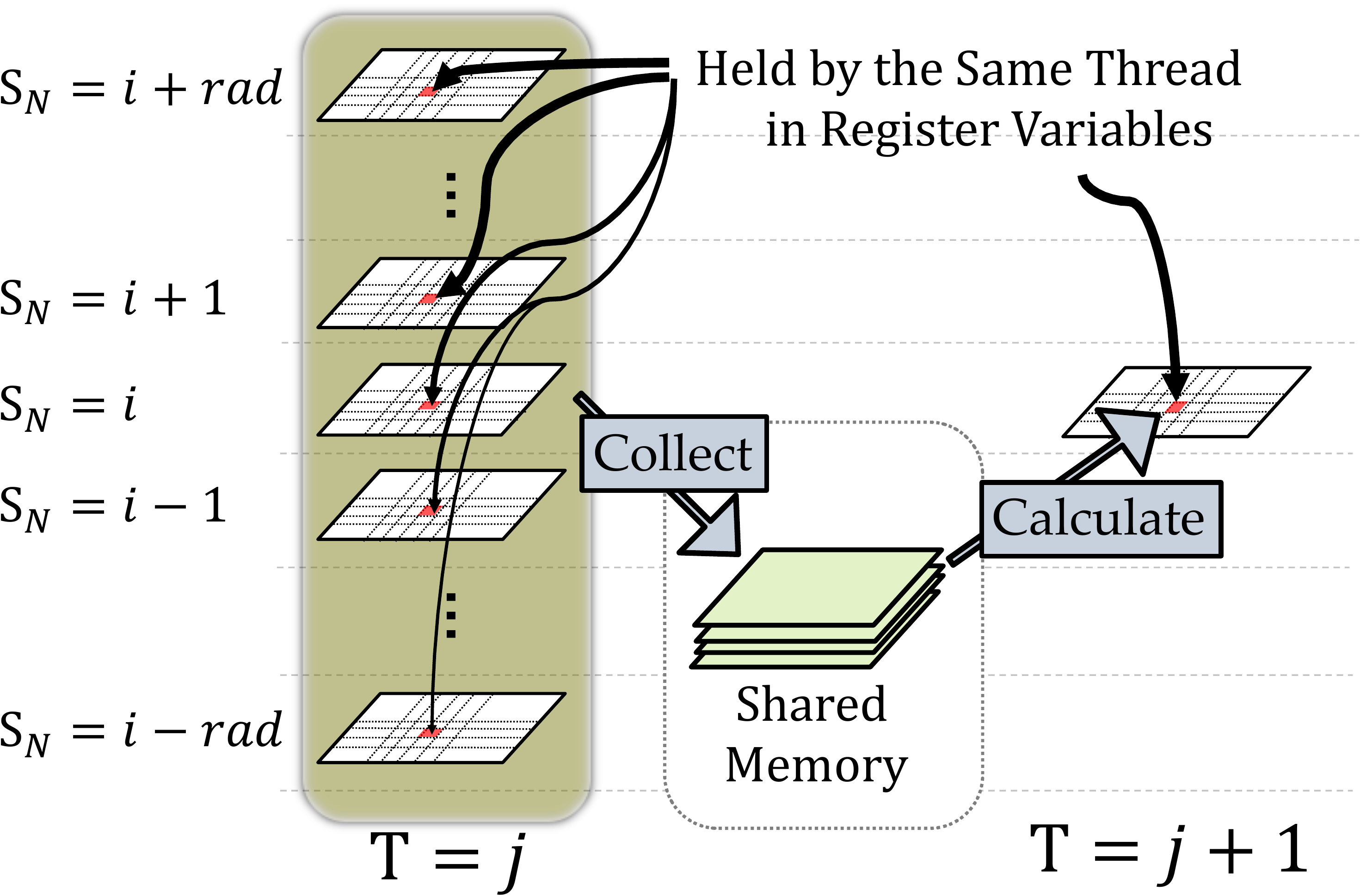}}\\
  \vspace{-0.18cm}
  \subfloat{\includegraphics[width=0.40\textwidth]{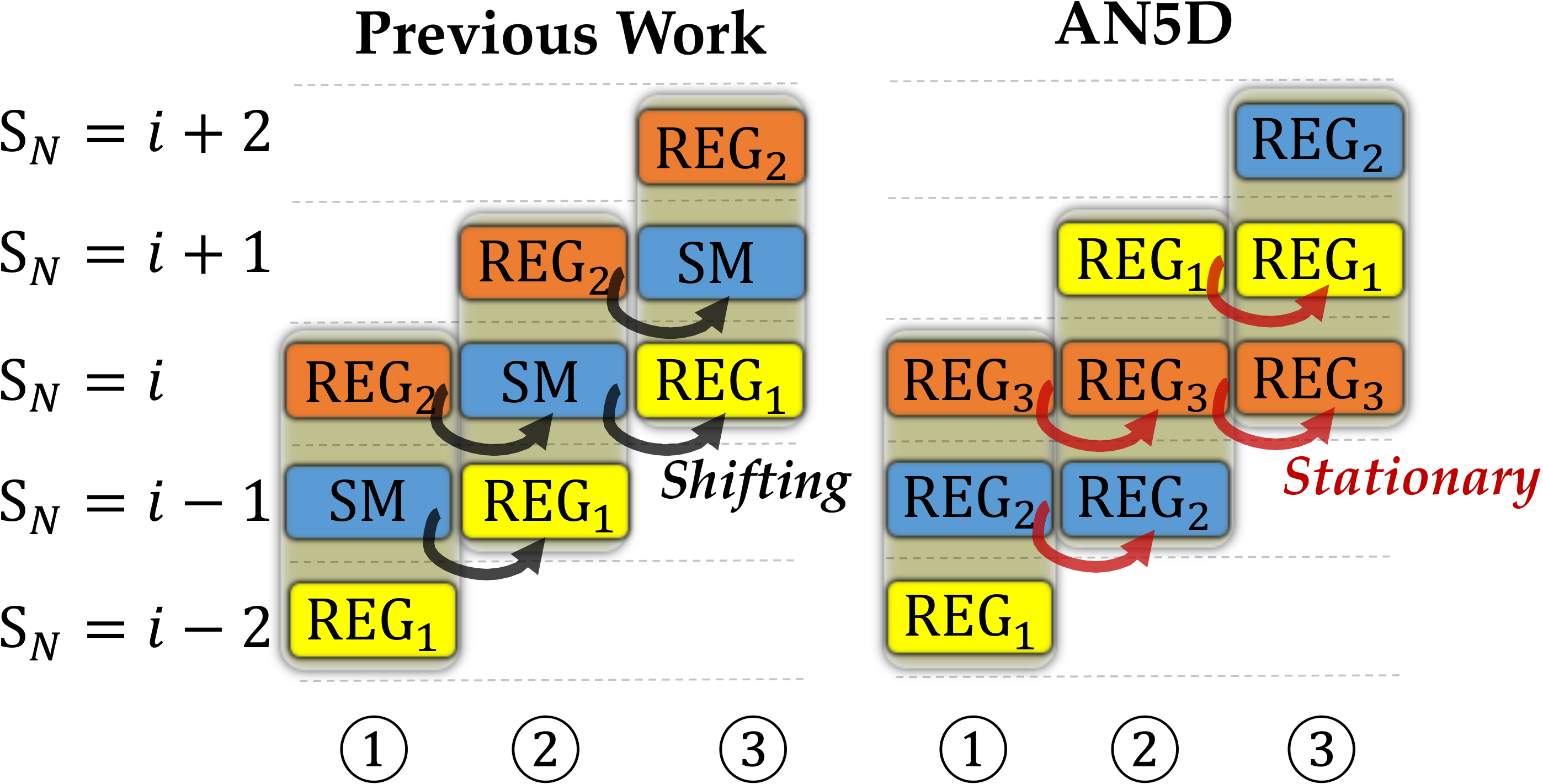}}
  \centering
  \vspace{-0.22cm}\hspace*{-1.5em}
  \caption{AN5D's on-chip memory management. (a) Shared memory use. (b) Register allocation.}
  \label{fig:memory-management}
\vspace{-1ex}  
\begin{tikzpicture}[remember picture,overlay,>=stealth]
  \node[font=\bfseries] at (-3.8,8.8) {\large (a)};
  \draw[line width=0.05mm] (-4.1,4.9) -- (4.1,4.9);
  \node[font=\bfseries] at (-3.8,4.55) {\large (b)};
\end{tikzpicture}
  \vspace*{-0.6cm}
\end{figure}

For neighboring accesses on each stencil calculation,
threads store data in shared memory as shown in Fig.~\ref{fig:memory-management}~(a).
While updating cells, both register variables and shared memory are accessed.
Cell values are fetched from registers
if the cells are owned by the requesting threads.
Otherwise, the accesses go to shared memory.
For diagonal-access free (star) stencils,
we eliminate the shared memory use in the upper ($\mathrm{S}_{N} = i + 1 \sim i + {rad}$) and lower sub-planes ($\mathrm{S}_{N} = i - {rad} \sim i - 1$).
In the case of associative box stencils, $1 + 2 \times {rad}$ consecutive sub-planes are simultaneously updated using values read from one sub-plane.
Based on the associativity, each sub-plane is computed through $1 + 2 \times {rad}$ partial summations. 

\vspace{-0.1cm}
\subsection{Optimizations}
To enable high degree temporal blocking, AN5D performs several optimizations to expose and efficiently utilize data locality at different levels of the memory hierarchy. To the authors knowledge, this is an unprecedented depth for temporal blocking in GPUs. It is important to note that GPUs are throughput-optimized processors that are commonly regarded as not ideal for the latency-sensitive deep temporal blocking approaches, in comparison to CPUs.
The following is list of the key optimizations that allows AN5D to achieve high performance that also scale to double-digit temporal steps in GPUs. 
\subsubsection{\bf Register Allocation}
Previous studies of N.5D blocking shift cells through registers (and shared memory) for holding a new sub-plane ~\cite{3.5d,rawatoptimizing,stencilgen1,stencilgen2}.
AN5D, however, uses a fixed set of registers for each sub-plane value by leveraging the fixed pattern of accesses.
Fig.~\ref{fig:memory-management}~(b) shows the difference between the register allocation of AN5D and previous work.
This optimization reduces data stores from $1 + 2 \times {rad}$ to $1$ for every sub-plane update,
leading to less data movement and register usage.

\subsubsection{\bf Double Buffering}
The shared memory use described in this section requires two thread-block synchronizations.
One is to wait for the result of the previous time-step,
and the other is to avoid overwriting shared memory while other threads are loading from it.
The latter synchronization can be skipped by utilizing an additional shared memory buffer to improve performance.
Although this increases shared memory requirements, shared memory usage is actually reduced compared to previous work~\cite{stencilgen1,stencilgen2}
for high degrees of temporal blocking since they use one separate shared memory buffer per combined time-step (see Table~\ref{tab:comp_to_stencilgen}).

\begin{figure}[b]
  \centering
  \begin{mdframed}[backgroundcolor=gray!12, linewidth=0.22mm,
      innertopmargin=0.05cm, innerbottommargin=0.025cm,
      innerleftmargin =0.25cm,
      leftmargin =0.02cm, rightmargin=0.02cm, usetwoside=false]
    \begin{lstlisting}
for (t = 0; t < I_T; t++)
 for (i = 1; i <= I_S2; i++)
  for (j = 1; j <= I_S1; j++)
    A[(t+1)%2][i][j] = (5.1f * A[t%2][i-1][j]
      + 12.1f * A[t%2][i][j-1] + 15.0f * A[t%2][i][j]
      + 12.2f * A[t%2][i][j+1] + 5.2f * A[t%2][i+1][j]) / 118;
  \end{lstlisting}
  \end{mdframed}
  \vspace{-0.3cm}
  \caption{\small j2d5pt code in C language}
  \label{fig:jacobi2d}
\end{figure}

\subsubsection{\bf Division of Streaming Dimension}
Depending on the ratio of input size to block size, there might not be enough thread-blocks participating in the computation to fully utilize the SMs.
To improve parallelism, AN5D supports dividing the streaming dimension and processing each stream block using a different thread-block,
at the cost of a minor amount of extra redundancy due to overlapping with previous and next stream blocks.
This increases the number of thread-blocks by the number of stream blocks as 
$n_{\mathrm{tb}}' = \left\lceil \nicefrac{{I}_{\mathrm{S}_{N}}}{h_{{\mathrm{S}_{N}}}} \right\rceil \times n_{\mathrm{tb}}$.
Here, $h_{\mathrm{S}_N}$ is the length of the divided stream blocks. The number of redundant sub-planes between the two consecutive stream blocks
is given by $2 \times \sum_{\mathrm{T}=0}^{b_{\mathrm{T}-1}} \left( {rad} \times (b_{\mathrm{T}} - \mathrm{T}) \right)$. The amount of overlapping
is variable for different $0 \leqslant T < b_{\mathrm{T}}$, and no overlapping is required for $T = b_{\mathrm{T}}$.

\subsection{Code Generation}
\subsubsection{\bf Host Code}

AN5D generates the host code in the form of repeated kernel calls.
Each kernel call performs one temporal blocking solution advancement of size $b_{\mathrm{T}}$.
Since our framework requires double-buffered stencil codes that use the modulo operator (t \% 2 $\rightarrow$ (t + 1) \% 2, (t + 1) \% 2 $\rightarrow$ t \% 2) as input (example shown in Fig.~\ref{fig:jacobi2d}),
the final result must be contained in the same global memory space to exactly follow the pattern of the original code.
Therefore, we adjust the final block of time-steps by reducing the degree of temporal blocking when $({I}_{\mathrm{T}}\ \mathrm{mod}\ b_{\mathrm{T}}) \neq 0$ or $(({I}_{\mathrm{T}} / b_{\mathrm{T}})\ \mathrm{mod}\ 2) \neq (b_{\mathrm{T}}\ \mathrm{mod}\ 2)$.
Since the size of the time dimension is not necessarily known at compile-time,
AN5D statically creates the conditional branches.

\begin{figure}[b]
  \centering
  \includegraphics[width=0.47\textwidth]{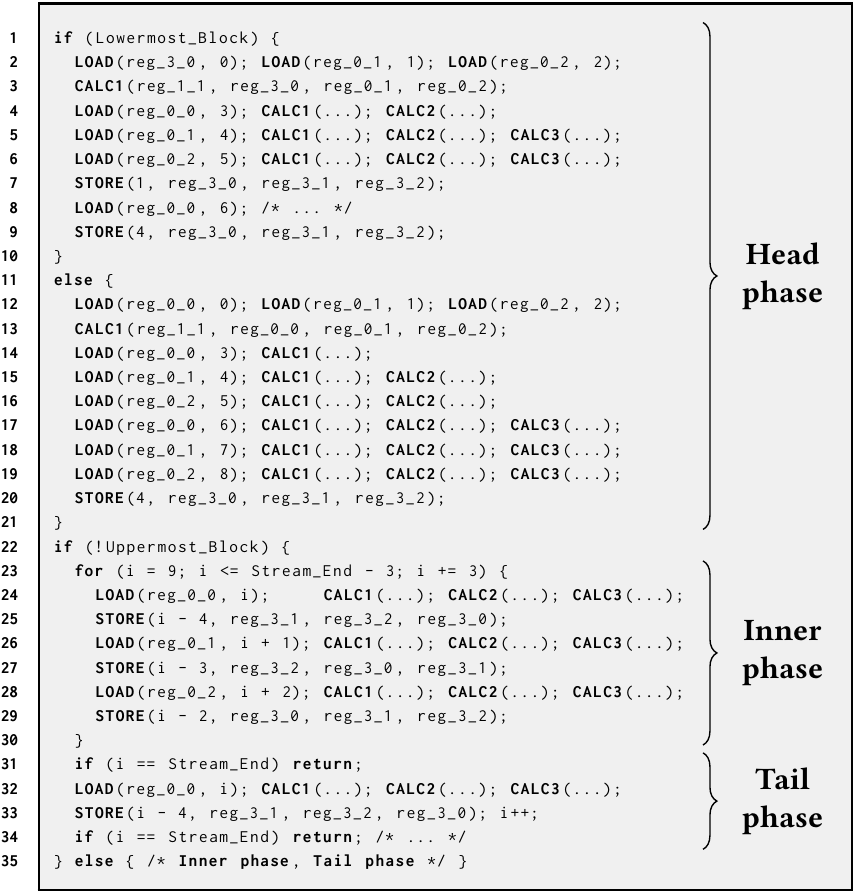}\vspace{-0.3cm}\\
  \caption{\small AN5D's generated code (j2d5pt, $b_{\mathrm{T}}=4$)}
  \label{fig:an5d-code}
\end{figure}

\subsubsection{\bf Kernel Code}
A kernel generated by AN5D consists of a sequence of macros, each of which
computes one sub-plane of a specific time-step. The macro calls take
the register variables that point to source and destination
sub-planes as input arguments, as well as the streaming index for global memory accesses.
Since the generated code performs no data shifting among the variables that are allocated
for sub-planes, the fixed register allocation depicted in Fig.~\ref{fig:memory-management}~(b)
is encoded as a sequence of macro arguments. The macros generated for each time-step conduct
its computation using the same double-buffered shared memory, and the result is written to the
destination register variable while avoiding writes to the halo region.

The generated sequence of macros implements the three phases of the computation:
\textit{head}, \textit{inner} and \textit{tail} (Fig.~\ref{fig:an5d-code}).
As described in Section~\ref{execmodel}, the head phase allocates registers
and computes the first $rad$ sub-planes.
Since control statements tend to increase register usage,
AN5D statically generates the code for this phase instead of using a loop (Lines~2-9 and 12-20 of Fig.~\ref{fig:an5d-code}).
For the inner phase, however, since the operations become repetitive,
a loop statement is used (Lines~23-30). Here, {\ttfamily N} and {\ttfamily M} of
{\ttfamily reg\_N\_M} represent $\mathrm{T}$ of Fig.~\ref{fig:tier} and the register ID of Fig.~\ref{fig:memory-management}~(b), respectively.
Each macro of {\ttfamily LOAD}, {\ttfamily CALC(1|2|3)} and {\ttfamily STORE}, loads, computes and stores one sub-plane of the time-step
$\mathrm{T}=0 \sim 4$, respectively, while the load, computation and store size is controlled by conditional branches.
The macro arguments are sequenced based on the value of $b_{\mathrm{T}}$ and which optimization (associative, diagonal-access free or none) is enabled.
Even though this loop statement can be unrolled,
our tests showed that doing so results in performance degradation due to increased instruction fetch latency.
The tail phase finishes the computation in a similar fashion to the head phase (Lines~31-34).

To reduce register pressure, we disable vectorized shared memory access since extra
registers are required to accommodate such accesses. To bypass automated vectorization
by NVCC, AN5D performs all shared memory loads through a device function that wraps
the shared memory access. In practice, disabling vectorized shared memory accesses
not only did not degrade performance, but in fact improved it due to the lowered register
pressure allowing higher thread-block-level parallelism. Moreover, AN5D assists
constant expansion by the CUDA compiler for double-buffered shared memory accesses
by inserting the explicit address or adjusting the number of times the buffer is
switched inside the loop statement.

\subsubsection{\bf Implementation}
AN5D generates the necessary CUDA code from an input C description. We use NVIDIA's CUDA compiler for compiling the generated code into an executable.
In our implementation, AN5D is integrated into a polyhedral compilation tool called PPCG~\cite{verdoolaege2013polyhedral, verdoolaege2012polyhedral}
that supports
\linebreak
CUDA code generation.
PPCG extracts the polyhedral representation~\cite{verdoolaege2010isl} from input (which is currently limited to C language) and this representation mainly consists of three parts which indicate the iteration domain, schedule and array accesses.
From these factors, PPCG computes various kinds of dependencies
and allows loop rescheduling including general loop tiling and hybrid tiling
on the polyhedral representation.
Since PPCG's backend is designed for general loop code and can not support
the specific forms AN5D requires,
we implement a dedicated backend within PPCG for AN5D.

Based on the representation normalized (dead-code eliminated and loop rescheduled) by PPCG's frontend,
our backend detects stencil patterns under the following rules:
\begin{itemize}[leftmargin=0.1in,rightmargin=0.05in, itemsep=2pt]
\item The statement describing array accesses is singleton and has only one store access.
Moreover, the addresses to read from the array are static.
\item All dimensions (time and space) are iterated by one loop, and multi-dimensional array addressing is used rather than linear addressing.
\item All the iterations for the spatial dimensions are data independent. Thus, the time loop is the outermost loop.
Moreover, the loop after the time loop represents the streaming dimension.
\end{itemize}
Although our implementation is only tested using inputs that follow the above restrictions,
we expect that existing techniques of polyhedral compilation can allow AN5D to accept a wider range of stencil codes.

Parameters such as $b_{\mathrm{T}}$, $b_{\mathrm{S}_i}$ and $h_{\mathrm{S}_N}$ are passed as compile-time parameters
to our backend, while input size (${I}_{\mathrm{S}_i}$) and time-step count (${I}_{\mathrm{T}}$) can be modified at run-time.
Optimizations such as diagonal-access free and associative stencil optimizations or disabling vectorized shared memory accesses
are enabled automatically by our framework but can be disabled using compile-time switches if necessary.
 \vspace*{-0.1cm}
\section{Performance Model}\label{model}

$b_{\mathrm{S}_i}$, $b_{\mathrm{T}}$ and $h_{\mathrm{S}_N}$ need to be tuned in our framework
for every given stencil and target hardware to maximize performance.
To prune the parameter search space and guide the performance tuning in our framework,
we construct a performance model based on the roofline method~\cite{roofline}.

First, we calculate the number of threads involved in the computation of any arbitrary
stencil supported by our framework, and classify the threads based on the operations
that they perform. Operations of importance for us are computation, global memory and shared memory accesses.
Based on this, we classify the threads into four categories: \textit{out-of-bound},
\textit{boundary}, \textit{redundant} and \textit{valid}. 
Out-of-bound threads are the ones that fall outside of the input grid space due to spatial blocking.
These threads still perform writes to shared memory to avoid extra branching in the kernels,
but avoid all global memory accesses and computation. Boundary threads are the ones
that load the cells holding the input boundary conditions located at the extremes
of the input grid. These threads perform shared memory reads and writes and global memory reads
but no global memory writes or computation. Redundant and valid threads are also defined as threads
that fall inside and outside of halo regions within the spatial blocks, respectively, both of which
perform shared memory read and write, global memory read, and computation, but only
the \textit{valid} threads perform global memory writes. Based on this classification and considering
variable halo size for $\mathrm{T}=0 \sim b_{\mathrm{T}}$ and extra overlapping caused by stream
blocking, we develop formulas to calculate the number of threads in each group based on
stencil shape, radius and input size, and the aforementioned parameters. We then calculate the total
number of threads that perform computation ($th_{\mathrm{comp}}$), shared memory read/write
($th_{\mathrm{sm\_read}}$/$th_{\mathrm{sm\_write}}$) and global memory read/write ($th_{\mathrm{gm\_read}}$/$th_{\mathrm{gm\_write}}$).

\begin{table}[b]
\centering
\renewcommand\tabcolsep{2pt}
\renewcommand{\arraystretch}{1}
\caption{\small Shared Memory Access per Thread}
\label{tab:sm_access}
\vspace{-0.35cm}
\resizebox{\columnwidth}{!}{%
\begin{tabular}{ c | c | c | c | c |}
\cline{2-5}
                                        & Shape & Read (Expected)                                                     & Read (Practical)                                   & Write \\
\hline
\multicolumn{1}{|c|}{\multirow{2}*{2D}} & Star  & $2 \times rad$                              & $2 \times rad$             & 1     \\
\cline{2-5}
\multicolumn{1}{|c|}{}                  & Box   & $(2 \times rad + 1)^2 - (2 \times rad + 1)$ & $(2 \times rad + 1) - 1$   & 1     \\
\hline
\multicolumn{1}{|c|}{\multirow{2}*{3D}} & Star  & $4 \times rad$                              & $4 \times rad$             & 1     \\
\cline{2-5}
\multicolumn{1}{|c|}{}                  & Box   & $(2 \times rad + 1)^3 - (2 \times rad + 1)$ & $(2 \times rad + 1)^2 - 1$ & 1     \\
\hline
\end{tabular}
}
\end{table}

In the next step, we need to determine how much computation or memory access is performed by each thread.
In practice, only one cell is read from global memory per thread when $\mathrm{T}=0$ and one cell is written when $\mathrm{T}=b_{\mathrm{T}}$.
Hence, the total global memory traffic ($total_{\mathrm{gm}}$) is $(th_{\mathrm{gm\_read}} + th_{\mathrm{gm\_write}}) \times n_{\mathrm{word}}$.

For shared memory traffic, we follow a similar approach. Table~\ref{tab:sm_access} shows the number
of shared memory reads and writes for threads that are involved in such operations. The number of
shared memory writes per thread is fixed to one (Section~\ref{execmodel}). For read, the numbers are obtained by deducting
the number of accesses that go to registers $(2 \times rad + 1)$ from the total number of cells
involved in the stencil computation. In practice, we noticed that the model underestimated performance
for box stencils when we used these values. Upon analyzing the generated PTX code we realized that NVCC was automatically
caching some of the shared memory data in registers, reducing the number of shared memory reads per
thread to one read per column in the stencil. Hence, we divide our expected shared memory reads per thread
by $(2 \times rad + 1)$ to get the practical value. Finally, we calculate the total shared memory traffic
($total_{\mathrm{sm}}$) similar to $total_{\mathrm{gm}}$.

For calculating the total number of floating-point operations,
we have to consider the equation associated with each stencil individually.
For synthetic star/box stencils in which the computation
is a straightforward dot product, all multiplications except the last one
are followed by an addition and hence, are merged into FMA operations.
However, for stencils such as Jacobi 2D (Fig.~\ref{fig:jacobi2d}),
since we use \texttt{{-}{-}use\_fast\_math} as a compiler switch,
we have to consider the alternative implementations of division and sqrt operations
enabled by this switch. Specifically, division is implemented as multiplication when
this switch is used. Profiling the number of floating-point operations using NVPROF
showed that in some cases, for stencils that use an equation similar to Jacobi 2D,
the compiler expands the statement inside the parenthesis over the division
and when the division is replaced with multiplication, the multiplications and additions
are merged into FMA operations. Based on these transformations, we determine
the mapping of each stencil's equation to ADD, MUL and FMA operations and
knowing the number of threads involved in the computation, we calculate the total
number of floating-point operations that are performed ($total_{\mathrm{comp}}$).

In the next step, we consider three possible bottleneck points: compute, global memory, and shared memory. We ignore registers as a bottleneck since we assume that, as long as no
register spilling occurs, the register bandwidth is high enough not to become a bottleneck.
We calculate the expected kernel run time for each level of bottleneck by dividing the total
traffic/computation involved for that level by its associated peak performance.
For compute, we use the theoretical peak compute performance of the devices ($peak_{\mathrm{comp}}$).
However, this peak performance is only valid if all the computation is mapped to FMA operations.
Hence, we calculate ALU utilization efficiency as
$eff_{\scriptscriptstyle{\mathrm{ALU}}} = \frac{2 \times op_{\scriptscriptstyle{\mathrm{FMA}}} + op_{\scriptscriptstyle{\mathrm{MUL}}} + op_{\scriptscriptstyle{\mathrm{ADD}}} + op_{\scriptscriptstyle{\mathrm{OTHER}}}}{2 \times (op_{\scriptscriptstyle{\mathrm{FMA}}} + op_{\scriptscriptstyle{\mathrm{MUL}}} + op_{\scriptscriptstyle{\mathrm{ADD}}} + op_{\scriptscriptstyle{\mathrm{OTHER}}})}$.

Then, we calculate run time assuming compute-bound operation as
$time_{\mathrm{comp}} = \frac{total_{\mathrm{comp}}}{peak_{\mathrm{comp}} \times eff_{\scriptscriptstyle{\mathrm{ALU}}}}$.
For shared and global memory, we use open-source benchmarks to measure practical peak performance ($peak_\mathrm{{sm|gm}}$) on the GPUs;
specifically, gpumembench~\cite{gpumembench} for measuring shared memory bandwidth (after adjusting the default vector size),
and BabelStream~\cite{babel} for measuring global memory bandwidth. Since the measured peak
performance for both memory types was different depending on data type (with the
difference being relatively large for shared memory), we used the associated peak values depending on the data type
used in the computation. Run times assuming global and shared memory bottleneck ($time_{\mathrm{gm|sm}}$) were then
calculated by dividing $total_{\mathrm{sm|gm}}$ by $peak_{\mathrm{sm|gm}}$.

One more point needs to be considered before we can calculate the expected run time: SM utilization efficiency ($eff_{\scriptscriptstyle{\mathrm{SM}}}$).
This value depends on how many thread-blocks are involved in the computation ($n'_{\mathrm{tb}}$) and how many SMs exist on the device
($n_{\scriptscriptstyle{\mathrm{SM}}}$). The number of thread-blocks that can simultaneously reside on each SM is limited by two factors:
the hardware limit of 2048 threads per SM, and the 64|96 KB shared memory size per SM on modern NVIDIA GPUs. The former factor limits the
number of concurrent thread-blocks per SM to $\nicefrac{2048}{n_{\mathrm{thr}}}$ while for diagonally-access free and associative stencils,
the latter factor imposes a limit of $\frac{\textrm{64KB (or 96KB)}}{2 \times n_{\mathrm{thr}} \times n_{\mathrm{word}}}$ thread-blocks (See Table~\ref{tab:comp_to_stencilgen}).
In practice, even for double-precision data, the former limit will be smaller and hence, we calculate SM utilization efficiency as
$eff_{\scriptscriptstyle{\mathrm{SM}}} = \nicefrac{\left\lfloor \frac{n'_{\mathrm{tb}}}{\nicefrac{2048}{n_{\mathrm{thr}}}} \right\rfloor}{\left\lceil \frac{n'_{\mathrm{tb}}}{\nicefrac{2048}{n_{\mathrm{thr}}}} \right\rceil}$. Finally, we calculate expected run time as
$time_{\mathrm{model}} = \frac{max(time_{\mathrm{comp}}, time_{\mathrm{sm}}, time_{\mathrm{gm}})}{eff_{\scriptscriptstyle{\mathrm{SM}}}}$.
 \vspace*{-0.1cm}
\section{Methodology}
\subsection{Benchmarks}

We evaluate our framework using a wide range of synthetic and general stencil benchmarks shown in Table~\ref{tab:benchmarks}.
In this table, \textit{c} denotes compile-time constant coefficients and
\textit{x} denotes the radius for the synthetic stencils. All other stencils have a
radius of one except j2d9pt which is a 2\textsuperscript{nd}-order stencil.
The synthetic benchmarks include all single-array single-statement box and star stencil shapes from 1st to 4th-order.
These benchmarks are specifically chosen to allow fair comparison with previous work;
some have a same-shaped equivalent among the synthetic benchmarks but
different computation pattern.

For each benchmark, we wrote C code and then generate the associated
CUDA host and kernel code for AN5D, general loop tiling, and hybrid tiling
from the same input code. To enable loop unrolling optimization for loop tiling and hybrid tiling,
we have to use static input sizes in the source code; this means that the source code needs to
be recompiled each time the input size needs to be changed which limits the usability of these
methods for real-world applications. However, our framework does not suffer from this limitation.
For STENCILGEN, we use the kernels released by the authors, available at https://github.com/\linebreak pssrawat/IEEE2017.
Comparison between frameworks is limited to stencil types available in this repository.

We use an input size of 16,384\textsuperscript{2} for 2D and 512\textsuperscript{3}
for 3D stencils with 1,000 iterations. These input sizes align with previous work~\cite{stencilgen1,Rawat:2018:ROS:3178487.3178500}
and are specifically chosen so that the GPUs are well-utilized and measured performance is stable,
without unnecessarily increasing the benchmark time. We also evaluate all benchmarks with both
single-precision and double-precision floating-point cell values. Each benchmark
is repeated five times (on top of an initial warm-up run) and the average performance
is reported (excluding PCI-E transfer time). Minimum run time for one instance of each benchmark is 400 ms.

\subsection{Hardware and Software}
We evaluate all benchmarks on the latest NVIDIA Tesla architectures: Pascal and Volta .
The specifications of these cards are shown in Table~\ref{tab:gpu_specs}. As for the external memory and shared memory bandwidth of the GPUs, we use open-source benchmarks to measure practical peak performance on the GPUs;
specifically, gpumembench~\cite{gpumembench} for measuring shared memory bandwidth (after adjusting the default vector size),
and BabelStream~\cite{babel} for measuring global memory bandwidth.

\begin{table}[b]
\small
\centering
\renewcommand{\arraystretch}{0.8}
\setlength\tabcolsep{3pt}
\newcolumntype{L}{ >{\raggedright\let\newline\\\arraybackslash} m }
\newcolumntype{M}{ >{\centering\let\newline\\\arraybackslash} m }
\caption{\small Benchmarks}
\label{tab:benchmarks}
\vspace{-0.35cm}
\footnotesize\begin{tabular}{ L{1.3cm} L{5.2cm} M{1.4cm}}
\toprule
  Stencil               & Computation                                                               & $\text{FLOP}/\text{Cell}$\\
\toprule
  star2d\{x\}r
  ${\mathrm{x}\in[1,4]}$  & $c_{(x,y)}f_{(x,y)} +\sum\limits_{i=-x,i\neq0}^{x} $ $(c_{(x+i,y)}f_{(x+i,y)} + c_{(x,y+i)}f_{(x,y+i)})$                   & $8 \mathrm{x} + 1$                  \\
\hline
  box2d\{x\}r
  ${\mathrm{x}\in[1,4]}$  & $\sum\limits_{i=-x}^{x} \sum\limits_{j=-x}^{x}$
                           $c_{(x+i,y+j)}f_{(x+i,y+j)}$                                          & ${2 \times}$ ${ (2 \mathrm{x} + 1)^2 - 1}$ \\
\hline
  j2d5pt                & $(c_{(x,y)}f_{(x,y)} + \sum\limits_{i=-1,i\neq0}^{1}$ $ (c_{(x+i,y)}f_{(x+i,y)} + c_{(x,y+i)}f_{(x,y+i)}))/c_{0}$                                        & 10                                  \\
\hline
  j2d9pt                & $(c_{(x,y)}f_{(x,y)} + \sum\limits_{i=-2,i\neq0}^{2} $ $ (c_{(x+i,y)}f_{(x+i,y)} + c_{(x,y+i)}f_{(x,y+i)}))/c_{0}$                                        & 18                                  \\
\hline
  \mbox{j2d9pt-gol}    & $(\sum\limits_{i=-1}^{1} \sum\limits_{j=-1}^{1}$ $c_{(x+i,y+j)}f_{(x+i,y+j)}) / c_{0}$                                                       & 18                                  \\
\hline
  gradient2d            & $c_{(x,y)}f_{(x,y)} + 1.0 / \mathrm{sqrt}(c_{0} + \sum\limits_{i=-1,i\neq0}^{1}$ $ ((f_{(x,y)}-f_{(x+i,y)}) \times (f_{(x,y)}-f_{(x+i,y)}) + 
					  (f_{(x,y)}-f_{(x,y+i)}) \times (f_{(x,y)}-f_{(x,y+i)})))$                          & 19                                  \\
\hline
  star3d\{x\}r
  ${\mathrm{x}\in[1,4]}$  & $c_{(x,y)}f_{(x,y)} + \sum\limits_{i=-x,i\neq0}^{x} $ $( c_{(x+i,y,z)}f_{(x+i,y,z)} + c_{(x,y+i,z)}f_{(x,y+i,z)} + c_{(x,y,z+i)}f_{(x,y,z+i)})$                      & ${12 \mathrm{x} + 1}$                 \\
\hline
  box3d\{x\}r
  ${\mathrm{x}\in[1,4]}$  & $\sum\limits_{i=-x}^{x} \sum\limits_{j=-x}^{x} \sum\limits_{k=-x}^{x} $ $c_{(x+i,y+j,z+k)}$ $f_{(x+i,y+j,z+k)}$                          & ${2 \times}$ ${(2 \mathrm{x} + 1)^3 - 1}$ \\
\hline
  j3d27pt               & $(\sum\limits_{i=-1}^{1} \sum\limits_{j=-1}^{1} \sum\limits_{k=-1}^{1} $ $c_{(x+i,y+j,z+k)} $ $f_{(x+i,y+j,z+k)}) / c_{0}$                                   & 54                                  \\
\bottomrule
\end{tabular}
\end{table}

\begin{table*}[t]
\begin{minipage}{\textwidth} 
\centering
\renewcommand\tabcolsep{2.9pt}
\newcolumntype{M}{ >{\centering\let\newline\\\arraybackslash} m }
\renewcommand{\arraystretch}{1}
\caption{\small GPU Specifications (Float | Double)}
\label{tab:gpu_specs}
\vspace*{-0.35cm}
\begin{tcolorbox}[colback=white,boxsep=0pt,left=2pt,right=2pt,top=1pt,bottom=1pt, arc=0mm, boxrule=\arrayrulewidth, colframe=black, enlarge top by=-0.1cm, enlarge bottom by=-0.1cm]
  \small\begin{tabular}{ M{2.6cm} | M{3cm} | M{2.9cm} | M{3.55cm} | M{3.4cm} | M{0.85cm} }
GPU                        &
                                                      Performance (GFLOP/s) & Peak External
										                            Memory Throughput (GB/s)  & Measured External
										                                                        Memory Throughput (GB/s)  & Measured Shared
										                                                                                    Memory Throughput (GB/s)  & SM Count\\
\hline
\multirow{1}*
{\shortstack{Tesla P100 SXM2}}       & 10,600  | 5,300              & 720 | 720                       & 535 | 540                      & \hspace{0.5em}9,700  | 10,150                   & \multirow{1}*{56}\\
\hline
\multirow{1}*
{\shortstack{Tesla V100 SXM2}}        & 15,700  | 7,850              & 900  | 900                      & 791 | 805                       & 10,650 | 12,750                   & \multirow{1}*{80}\\
\end{tabular}
\end{tcolorbox}
\end{minipage}
\end{table*}

\begin{figure*}[t]
  \centering
  \vspace{-0.7cm}
  \subfloat{\includegraphics[width=0.495\textwidth]{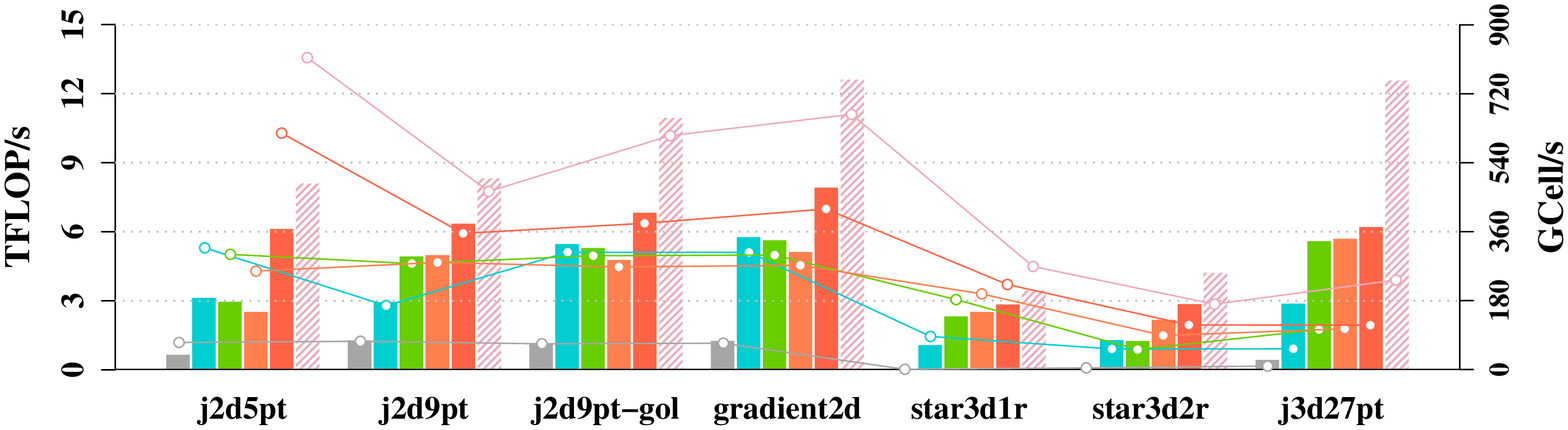}}
  \hfill
  \subfloat{\includegraphics[width=0.495\textwidth]{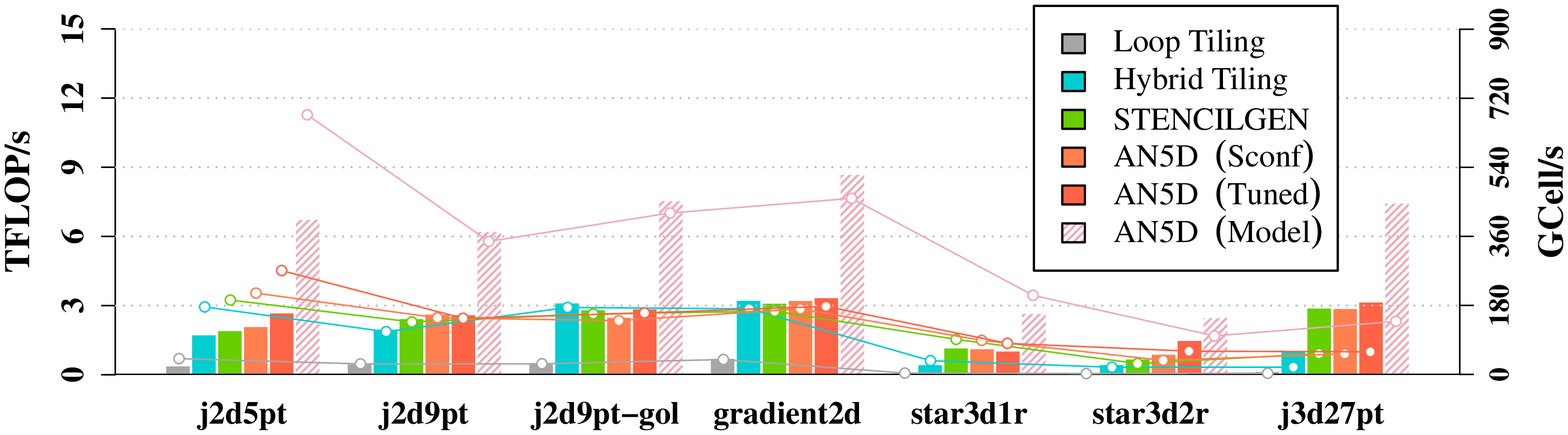}}\\
  \vspace{-0.55cm}
  \centering
  \subfloat{\includegraphics[trim=0 0 0 0.2cm,clip,width=0.495\textwidth]{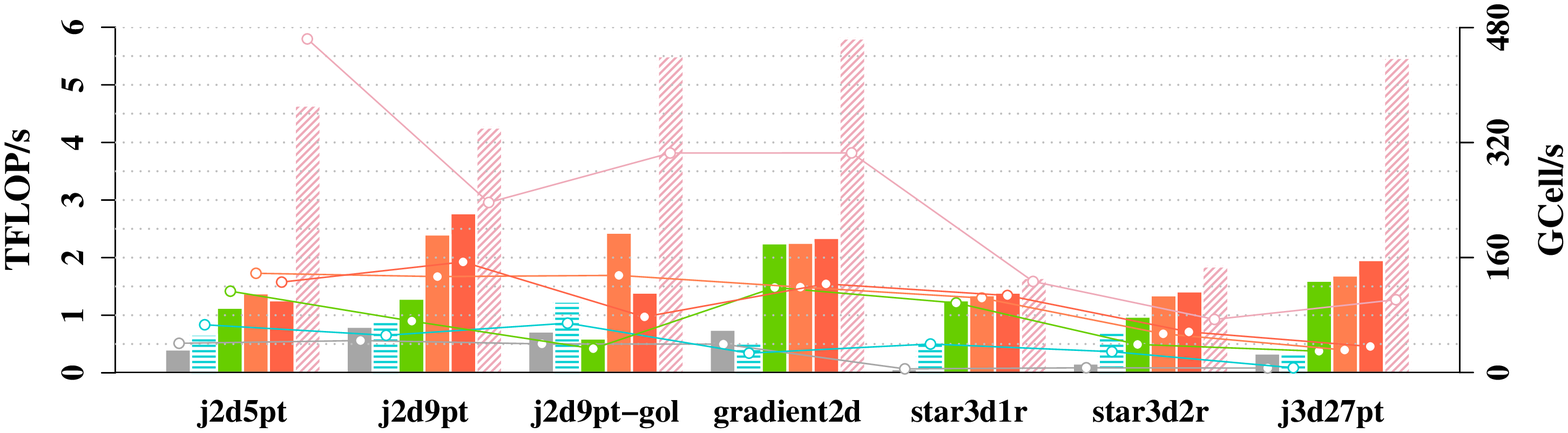}}
  \hfill
  \subfloat{\includegraphics[trim=0 0 0 0.2cm,clip,width=0.495\textwidth]{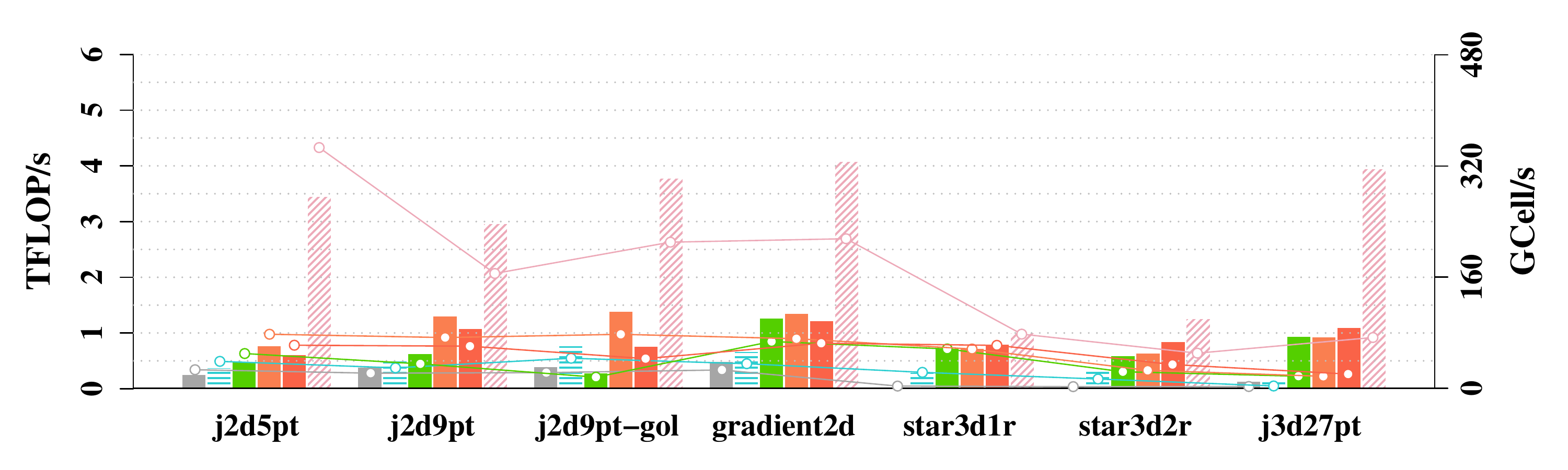}}\\
  \centering
  \vspace{-0.4cm}
  \caption{Performance on Tesla V100 (left) and P100 (right) with float (top) and double (bottom) data types}
  \label{fig:comp}
  \vspace{-0.1cm}
\end{figure*}

Both of our machines use CentOS 7.6 and Intel Xeon processors. For compiling the kernel codes,
we use CUDA 10.0.130 and associated NVIDIA driver v410.48. We compile all kernels using the following
set of compiler arguments: \texttt{"-gencode=arch=compute\_(60|70),code=sm\_(60|70) --use\_fast\_math -Xcompiler -O3 -fopenmp"}.
Even though \texttt{--use\_fast\_math} can reduce the numerical accuracy of the results, we use this switch
so that our evaluation aligns with previous work~\cite{stencilgen1,Rawat:2018:ROS:3178487.3178500}.
Moreover, for cases where complex mathematical operations are involved (sqrt and division),
not using this switch might make the benchmark compute-bound and eliminate the need
for shared memory and register-based optimizations.

\subsection{Parameter Tuning}
\label{tuning}

We measure the performance of AN5D using two configurations.
The first configuration called \texttt{Sconf} uses the same
kernel parameters as STENCILGEN: $b_{\mathrm{T}}=4$, $h_{\mathrm{S}_N}=128$
and $b_{\mathrm{S}_i}=\{32|128\}$ (2D|3D, respectively).
We also disable associative stencil optimization for 2D stencils
and streaming division for 3D stencils in this configuration since STENCILGEN does not use this optimization for such stencils.
The second configuration is called \texttt{Tuned}. For this configuration,
we first use our model to predict the performance for all valid parameter sets.
Specifically, we use $b_{\mathrm{T}} \in [1,16]$ for 2D,
and $b_{\mathrm{T}} \in [1,8]$ for 3D stencils, respectively. $b_{\mathrm{S}_i}$
for 2D stencils is chosen from the set of $\{128,256,512\}$, and for 3D, is chosen
from \{16$\scriptstyle\times$16, 32$\scriptstyle\times$16, 32$\scriptstyle\times$32, 64$\scriptstyle\times$16\}.
$h_{\mathrm{S}_N}$ is also chosen from the sets of $\{256,512,1024\}$ and $\{128,256\}$ for 2D and 3D
stencils, respectively. These settings result in 144 configurations for 2D,
and 64 configurations for 3D stencils per GPU, all of which can be searched in
a few seconds using our model. However, we experimentally find that
a minimum of $b_{\mathrm{T}}$$\times$$(2 \times rad + 1)$$+ b_{\mathrm{T}} + 20$ and
$2 \times b_{\mathrm{T}} \times (2 \times rad + 1) + b_{\mathrm{T}} + 30$ registers are used per thread,
for single and double-precision data types, respectively. Hence, we use these limits to prune
configurations which are expected to require more than the hardware limits of 255 registers per thread
or 65,536 registers per SM. Then, we sort the parameter sets based on the performance predicted by our model
and choose the top 5 for each GPU. Finally, we run these configurations on the GPUs and choose the one that
achieves the highest measured performance in each case.

For loop tiling, we use the default tiling size that is provided by PPCG.
Hybrid tiling provides parameters to optimize tile sizes along each time/spatial dimension,
and thread-block sizes along each spatial dimension.
We conduct a large-scale parameter search to find the optimal parameters
for each combination of stencil pattern and GPU.
Here, around 10,000 and 5,000 parameter configurations are explored for
each 2D ($b_\mathrm{T} = [2, 20]$, $b_\mathrm{S} = [1, 32]\times[32, 2048]$,
$n_\mathrm{thr}=[1, 32]\times[32, 1024]$) and 3D stencil ($[2, 12]$, $[1, 4]\times[1, 32]\times[32, 256]$,
$[1, 4]\times[1, 32]\times[32, 256]$), respectively. We set 8,192\textsuperscript{2} and 512\textsuperscript{3}
as 2D/3D grid size and 120 as iteration count for parameter search.
We conduct the parameter search with single-precision
and reuse the optimal parameters of each stencil for double-precision.
Even though this might not necessarily result in the best performance for the double-precision case,
the method would be the same as STENCILGEN and the \texttt{Sconf} configuration for AN5D where
the same configurations are used for single and double-precision data types.
To speed-up parameter tuning for the Tesla V100 GPU, we utilize the ABCI supercomputer environment
with CUDA 9.2, while parameter tuning for Tesla P100 and the actual performance measurement for both GPUs are done using the same local machine
as we use for other frameworks with CUDA 10.0.130. 

One final parameter to tune is the number of registers allocated per thread.
This value can be restricted using the NVCC option \texttt{{-}maxrregcount}.
Limiting register usage can allow more thread-blocks to reside on the same SM
at the same time, leading to higher parallelism and performance.
However, this restriction can also lead to register spilling
which would adversely affect performance. We encountered multiple
cases where limiting register usage per thread reduced register usage
\textit{without} spilling and consequently, increased performance
due to better SM utilization. Hence, for all benchmarks and all frameworks,
apart from the standard compilation where no register limit was imposed,
we also generated binaries with limits of 32 and 64 registers per thread, and chose
the best performance for each case. For the \texttt{Tuned} configuration of AN5D, we further added the limit of 96
register per thread since it proved to be useful for high-order stencils and high-degree temporal blocking.
 \vspace*{-0.2cm}
\section{Results}
\subsection{Comparison}
Fig.~\ref{fig:comp} shows performance comparison results for multiple stencils.
\texttt{Model} is the performance predicted by our model for the \texttt{Tuned}
configuration. On Tesla V100, taking both \texttt{Sconf} and \texttt{Tuned} results
into account, AN5D achieves the highest performance for both single and double-precision.
On Tesla P100, AN5D achieves the highest performance
except in the cases of j2d9pt-gol and star3d1r in single-precision
where hybrid tiling and STENCILGEN are faster than AN5D by 8\% and 3\%, respectively.
These two cases will be further discussed later in this section.

\begin{figure}[b]
  \centering
  \includegraphics[width=0.98\linewidth]{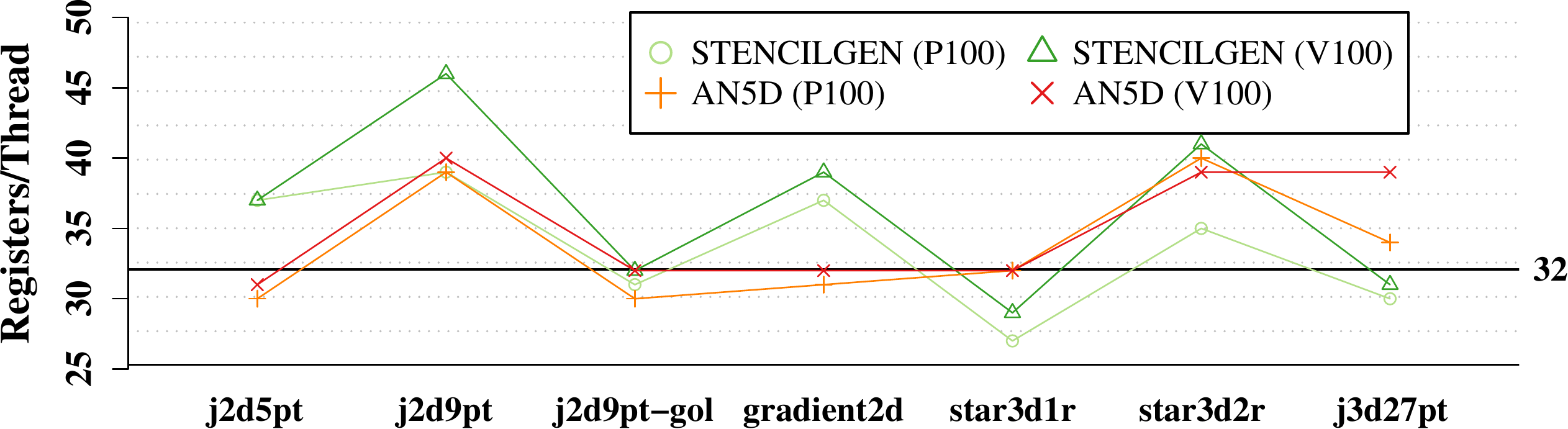}
  \vspace{-0.2cm}
  \caption{Register usage with no register limitation (float)}
  \label{fig:regcomp}
\end{figure}

Using the same configuration as STENCILGEN, AN5D improves performance for most cases and specifically exhibits
large performance improvements of up to 2x for double-precision benchmarks due to lower register pressure than STENCILGEN.
This shows that even though the focus of our framework is to enable high-degree temporal blocking,
it can still compete with or outperform state-of-the-art using same configurations.
Fig.~\ref{fig:regcomp} shows the register usage of STENCILGEN and \texttt{Sconf} configuration for AN5D for multiple stencils.
Even though in theory, AN5D requires $b_{\mathrm{T}}$ extra registers per thread for sub-plane management compared to STENCILGEN,
in practice it uses fewer registers on average.
Moreover, when we limit register usage per thread to 32 (maximum value to achieve 100\% SM occupancy),
none of the seven binaries generated by AN5D cause register spilling,
while STENCILGEN causes spilling for the second-order stencils (j2d9pt and star3d2r).

\begin{table*}[htbp]
\small
  \begin{minipage}{\textwidth} 
  \centering
  \caption{\small AN5D Configuration and Performance (Regs: Optimal register limitation per thread, ``-'': no limitation, Tuned \& Model: GFLOP/s)}
  \label{tab:confg_performance}
\vspace{-0.35cm}
\def\arraystretch{1}
\setlength\tabcolsep{2.4pt}
\scriptsize
\hspace*{-0.4em}
\begin{tcolorbox}[colback=white,boxsep=0pt,left=2pt,right=2pt,top=1pt,bottom=1pt, arc=0mm, boxrule=\arrayrulewidth, colframe=black, enlarge top by=-0.1cm, enlarge bottom by=-0.1cm]
  \begin{tabular}{l|cccccc|cccccc|cccccc|cccccc}
    \multicolumn{1}{c|}{\multirow{2}{*}{Pattern}} & \multicolumn{6}{c|}{\textbf{Tesla V100 (float)}} & \multicolumn{6}{c|}{\textbf{Tesla V100 (double)}}& \multicolumn{6}{c|}{\textbf{Tesla P100 (float)}} & \multicolumn{6}{c}{\textbf{Tesla P100 (double)}} \\
    & $b_{\mathrm{T}}$ & $b_{\mathrm{S}}$ & $h_{\mathrm{S}_N}$ & Regs & Tuned & Model
    & $b_{\mathrm{T}}$ & $b_{\mathrm{S}}$ & $h_{\mathrm{S}_N}$ & Regs & Tuned & Model
    & $b_{\mathrm{T}}$ & $b_{\mathrm{S}}$ & $h_{\mathrm{S}_N}$ & Regs & Tuned & Model
    & $b_{\mathrm{T}}$ & $b_{\mathrm{S}}$ & $h_{\mathrm{S}_N}$ & Regs & Tuned & Model
    \\
\hline
\rowcolor{lightgrey}
star2d1r & 10 & 256 & 256 & 64 & 5,631 & 7,330 &
10 & 256 & 256 & - & 3,306 &  4,177 &
13 & 512 & 256 & - & 2,507 &  6,091 &
15 & 128 & 1024 & - & 1,588 & 3,118
\\ 
star2d2r & 10 & 512 & 256 & 64 & 6,319 & 8,172 &
6 & 512 & 128 & 64 & 3,071 &  4,431 &
10 & 512 & 512 & - & 2,576 &  7,698 &
8 & 512 & 512 & 96 & 1,397 & 4,042
\\ 
\rowcolor{lightgrey}
star2d3r & 7 & 512 & 256 & 64 & 7,132 & 8,627 &
6 & 256 & 128 & 96 & 3,221 &  4,707 &
6 & 512 & 512 & 96 & 3,424 &  8,144 &
7 & 256 & 256 & - & 1,912 & 3,857
\\ 
star2d4r & 5 & 512 & 256 & - & 7,244 & 8,954 &
4 & 256 & 128 & 96 & 3,422 &  4,680 &
5 & 512 & 512 & - & 3,573 &  8,405 &
5 & 512 & 512 & - & 1,956 & 4,397
\\
\hdashline[1pt/2pt]
\rowcolor{lightgrey}
bo$\scriptstyle\times$2d1r & 10 & 256 & 256 & 96 & 6,693 & 11,327 &
10 & 256 & 256 & - & 2,984 &  5,664 &
10 & 256 & 512 & 64 & 2,823 &  7,804 &
8 & 128 & 128 & 96 & 1,959 & 3,660
\\ 
bo$\scriptstyle\times$2d2r & 5 & 256 & 256 & 64 & 9,163 & 12,473 &
3 & 256 & 128 & 64 & 4,686 &  5,858 &
5 & 256 & 512 & 64 & 4,626 &  8,578 &
5 & 256 & 512 & - & 2,673 & 4,289
\\ 
\rowcolor{lightgrey}
bo$\scriptstyle\times$2d3r & 2 & 256 & 128 & 96 & 10,227 & 12,391 &
2 & 256 & 128 & 64 & 5,507 &  6,196 &
2 & 256 & 128 & 96 & 5,598 &  8,584 &
2 & 128 & 128 & 96 & 3,652 & 4,244
\\ 
bo$\scriptstyle\times$2d4r & 4 & 512 & 256 & 96 & 10,772 & 13,241 &
1 & 256 & 128 & 96 & 5,770 &  6,576 &
4 & 512 & 512 & 96 & 6,546 &  9,174 &
1 & 128 & 128 & 96 & 3,921 & 4,556
\\
\hdashline[1pt/2pt]
\rowcolor{lightgrey}
j2d5pt & 10 & 256 & 256 & 64 & 6,160 & 8,144 &
10 & 256 & 256 & 96 & 1,258 &  4,642 &
13 & 512 & 256 & - & 2,708 &  6,768 &
15 & 128 & 1024 & - & 621 & 3,465
\\ 
j2d9pt & 5 & 256 & 256 & - & 6,398 & 8,370 & 
5 & 256 & 256 & 64 & 2,770 &  4,259 &
10 & 512 & 256 & 64 & 2,635 & 6,244 &
6 & 512 & 128 & 64 & 1,093 & 2,976
\\ 
\rowcolor{lightgrey}
j2d9pt-gol & 10 & 256 & 256 & - & 6,865 & 10,994 &
10 & 256 & 256 & - & 1,394 &  5,497 &
10 & 256 & 512 & 64 & 2,883 &  7,575 &
10 & 256 & 512 & - & 770 & 3,787
\\ 
gradient2d & 10 & 256 & 256 & 96 & 7,965 & 12,660 &
8 & 256 & 128 & 64 &  2,343 & 5,806 &
10 & 256 & 512 & 64 & 3,369 & 8,723 & 
8 & 128 & 128 & 96 & 1,234 & 4,091
\\
\hdashline[1pt/2pt]
\rowcolor{lightgrey}
star3d1r & 4 & 32$\scriptstyle\times$32 & 128 & 96 & 2,887 & 3,498 & 
4 & 64$\scriptstyle\times$16 & 128 & 32 & 1,393 & 1,647 & 
5 & 32$\scriptstyle\times$32 & 128 & 96 & 1,055 & 2,682 & 
3 & 32$\scriptstyle\times$32 & 128 & 32 & 805 & 1,015
\\ 
star3d2r & 3 & 32$\scriptstyle\times$32 & 128 & 32 & 2,910 & 4,268 & 
2 & 32$\scriptstyle\times$32 & 128 & 64 & 1,413 & 1,847 & 
2 & 32$\scriptstyle\times$32 & 128 & 96 & 1,545 & 2,512 & 
2 & 32$\scriptstyle\times$32 & 128 & 32 & 859 & 1,268
\\ 
\rowcolor{lightgrey}
star3d3r & 2 & 32$\scriptstyle\times$32 & 128 & 32 & 3,118 &4,518 & 
2 & 32$\scriptstyle\times$32 & 256 & 96 & 1,591 & 2,193 & 
2 & 32$\scriptstyle\times$32 & 128 & 32 & 1,523 & 3,117 & 
2 & 32$\scriptstyle\times$32 & 256 & - & 966 & 1,528
\\ 
star3d4r & 2 & 32$\scriptstyle\times$32 & 256 & 32 & 2,808 &4,063 & 
1 & 32$\scriptstyle\times$32 & 128 & 96 & 1,684 & 2,087 & 
1 & 64$\scriptstyle\times$16 & 128 & - & 1,656 & 2,824 & 
1 & 32$\scriptstyle\times$32 & 256 & 64 & 1,135 & 1,354
\\
\hdashline[1pt/2pt]
\rowcolor{lightgrey}
bo$\scriptstyle\times$3d1r & 3 & 32$\scriptstyle\times$32 & 256 & 32 & 6,284 &12,811 & 
3 & 32$\scriptstyle\times$16 & 128 & - & 2,888 & 5,552 & 
3 & 32$\scriptstyle\times$32 & 256 & 64 & 3,168 & 7,590 & 
3 & 32$\scriptstyle\times$32 & 128 & - & 1,671 & 4,015
\\ 
bo$\scriptstyle\times$3d2r & 1 & 32$\scriptstyle\times$16 & 128 & 96 & 8,666 &13,640 & 
1 & 32$\scriptstyle\times$16 & 128 & 96 & 5,024 & 6,820 & 
1 & 32$\scriptstyle\times$16 & 128 & 64 & 5,528 & 9,482 & 
1 & 32$\scriptstyle\times$16 & 128 & 96 & 3,189 & 4,741
\\ 
\rowcolor{lightgrey}
bo$\scriptstyle\times$3d3r & 1 & 64$\scriptstyle\times$16 & 128 & 96 & 9,351 &13,931 & 
1 & 32$\scriptstyle\times$16 & 128 & - & 2,993 & 7,599 & 
1 & 32$\scriptstyle\times$16 & 128 & 96 & 6,401 & 9,749 & 
1 & 32$\scriptstyle\times$16 & 128 & - & 1,934 & 4,874
\\ 
bo$\scriptstyle\times$3d4r & 1 & 64$\scriptstyle\times$16 & 256 & - & 9,707 & 15,248 & 
1 & 64$\scriptstyle\times$16 & 256 & - & 4,635 & 7,624 & 
1 & 32$\scriptstyle\times$16 & 128 & - & 3,056 & 9,928 & 
1 & 16$\scriptstyle\times$16 & 256 & - & 794 & 4,891
\\
\hdashline[1pt/2pt]
\rowcolor{lightgrey}
j3d27pt & 3 & 32$\scriptstyle\times$32 & 256 & 32 & 6,251 & 12,617 & 
3 & 32$\scriptstyle\times$16 & 128 & 64 & 1,957 & 5,468 & 
3 & 32$\scriptstyle\times$32 & 256 & 96 & 3,183 & 7,476 &
3 & 32$\scriptstyle\times$32 & 128 & 64 & 1,112 & 3,954
  \end{tabular}
\end{tcolorbox}  
  \end{minipage}
\end{table*}

Loop tiling fails to compete with any of the evaluated frameworks.
Hybrid tiling, on the other hand, achieves competitive performance for 2D stencils
but only outperforms AN5D in the case of j2d9pt-gol on P100. Based on profiling
results, hybrid blocking achieves better global memory performance than AN5D in this case,
leading to slightly higher performance.
For 3D stencils, this tiling method falls short of STENCILGEN and AN5D since,
as discussed in Section~\ref{related}, it is limited to smaller block sizes due
to lack of dimension streaming.

\begin{figure}[b]
  \centering
  \includegraphics[width=0.48\textwidth]{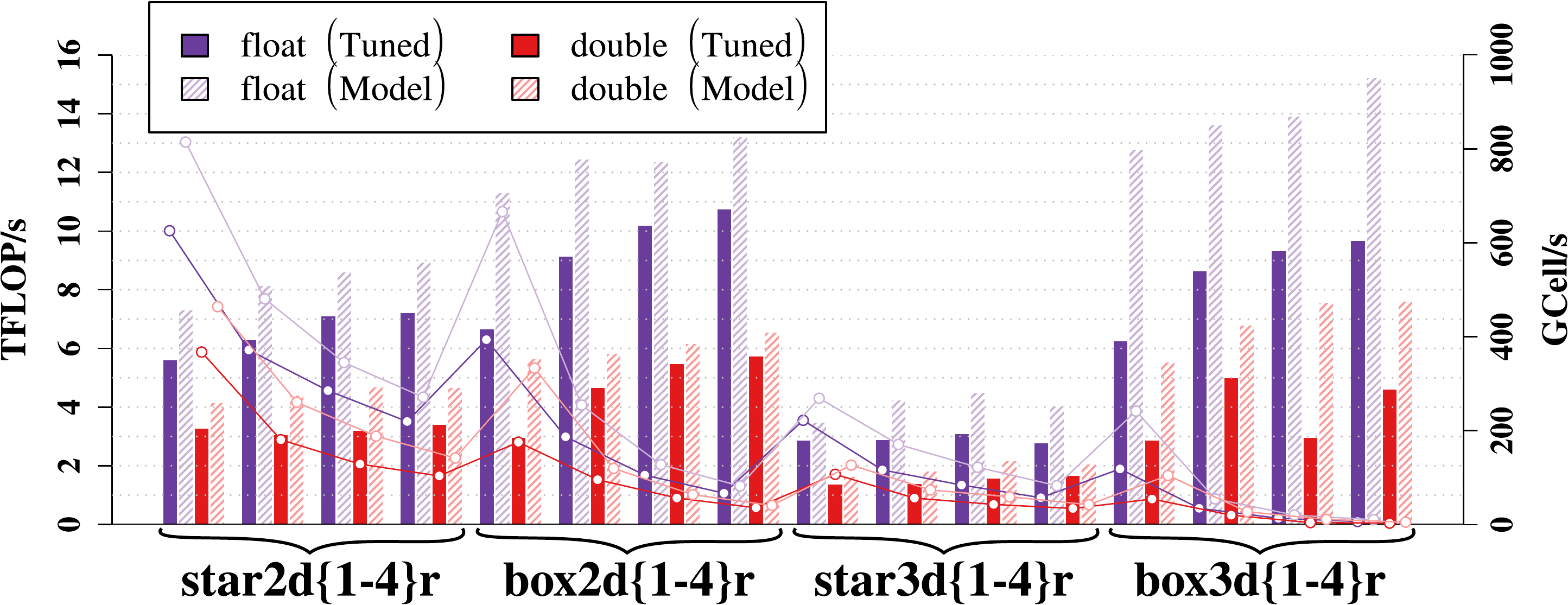}
  \vspace{-0.7cm}
  \caption{Performance of Star/Box Stencils on Tesla V100}
  \label{fig:star-box}
  \vspace{-0.3cm}
\end{figure}

Comparing the \texttt{Tuned} and \texttt{Sconf} configurations for AN5D, 
the \texttt{Tuned} configuration increases the performance in every case for single-precision benchmarks on Tesla V100.
For double-precision, however, performance degradation is observed for multiple stencils on both devices.
Further analysis showed that the CUDA compiler generates inefficient machine code for stencils
which use double-precision division (j$\ast$$\ast$$\ast$$\ast$ stencils in Fig.~\ref{fig:comp}),
resulting in noticeable slow-down compared to same-shaped stencils
that do not use this operation (Fig.~\ref{fig:star-box}). Since our model is unaware of this problem,
it fails to predict the best configuration for such stencils. We experimentally discovered that replacing the
double-precision division "$/N$" with "$\times(1/N)$" can be used as a work-around here,
but we did not use it in our evaluation to keep the comparison fair.

\subsection{Model Accuracy}
Table~\ref{tab:confg_performance} shows the best-performing AN5D configuration among top 5 configurations predicted by our model for all our evaluated stencils,
alongside with optimal register per thread limit (Regs), and measured (Tuned) and predicted (Model) performance for this configuration.
We define model accuracy as the ratio of \texttt{Tuned} to \texttt{Model}.
Our model exhibits an average accuracy of 49\% (16$\sim$86\%) on P100 and 67\% (25$\sim$89\%) on V100 when all cases
from Table~\ref{tab:confg_performance} are considered. Excluding the benchmarks that use the division operation and achieve lower-than-expected
performance with double-precision data, the model accuracy improves to 53\% (16$\sim$86\%) and 71\% (39$\sim$89\%), on P100 and V100, respectively.
The lowest accuracy values are obtained for box3d3r and box3d4r where register usage that is not considered in our model becomes a bottleneck,
especially on the P100 device. Since our model predicts shared memory as the performance bottleneck in every case except box3d3r and box3d4r,
the model accuracy can be considered as an estimation for shared memory efficiency on these GPUs.
Profiling multiple of our benchmarks on P100 with the same kernel that was used on V100 showed that P100 achieves less than half the shared memory
bandwidth of V100 for the same kernels, despite the fact that the difference between the measured shared memory bandwidth of these devices
is less than 10\% (Table~\ref{tab:gpu_specs}). Hence, we can conclude that Tesla V100 is more suitable for N.5D blocking as it achieves higher shared
memory efficiency, pointing to a more efficient shared memory architecture and controller on this device compared to Tesla P100.
One negative side-effect of the lower-than-expected shared memory efficiency of P100 is that our model tends to overestimate the optimal
degree of temporal blocking on this device. For example, for the star3d1r benchmark on P100 which is the only case we report lower performance than STENCILGEN,
if we use the exact same configuration as reported in Table~\ref{tab:confg_performance} but reduce $b_{\mathrm{T}}$ to 3,
performance will increase to 1,263 GFLOP/s which is higher than STENCILGEN, enabling us to achieve higher performance than this state-of-the-art framework
for every benchmark and device. It is likely that if we consider the shared memory efficiency of the devices in our model, we can find better configurations also for other benchmarks.

\begin{figure}[b]
  \vspace{-0.4cm}
  \hspace*{-0.15cm}
  \centering
  \subfloat{\includegraphics[trim=0.4cm 0.2cm 0.2cm 0.2cm,clip,width=0.28\textwidth]{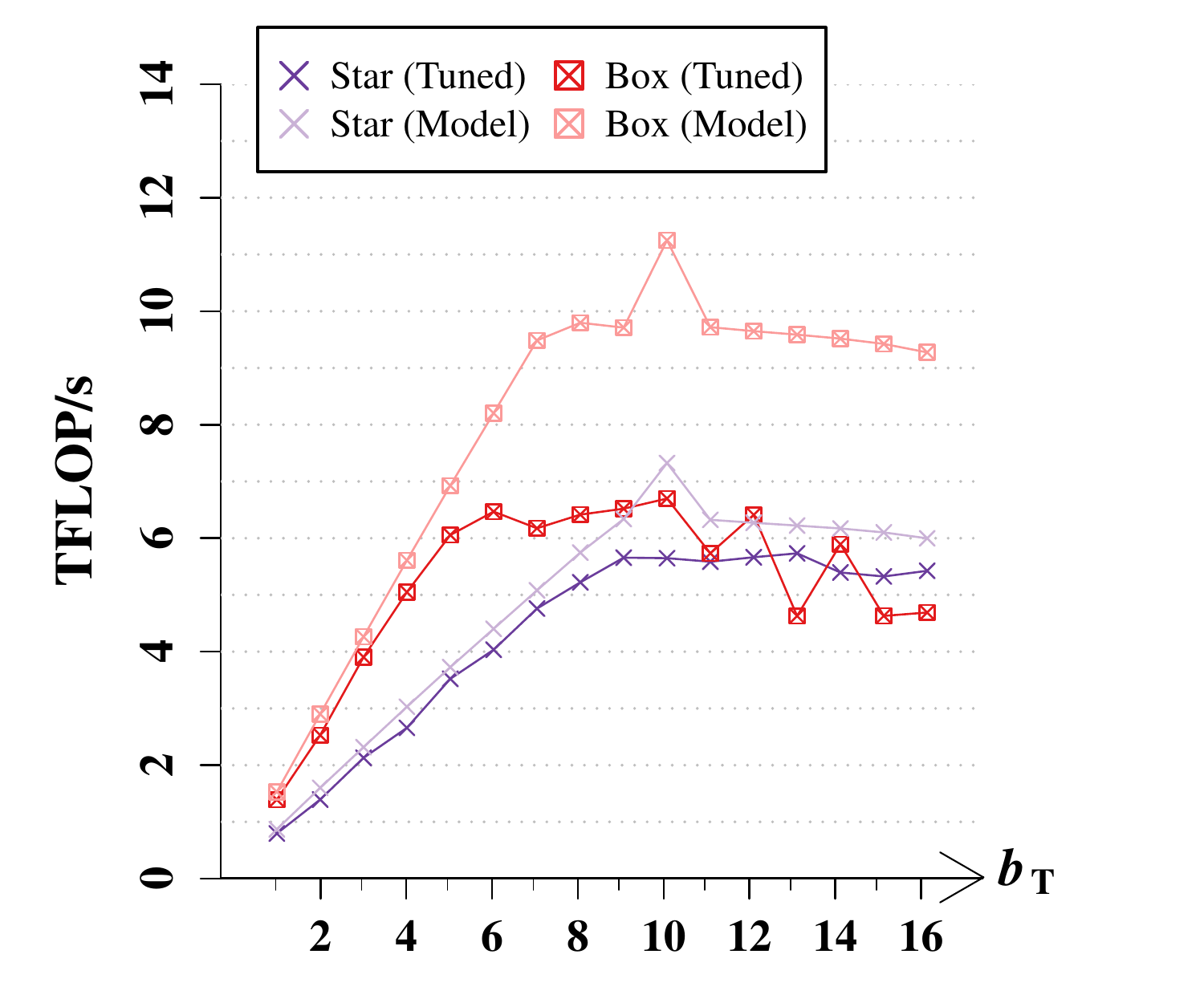}}
  \hspace*{-0.8cm}
  \subfloat{\includegraphics[trim=0.4cm 0.2cm 0.2cm 0.2cm,clip,width=0.28\textwidth]{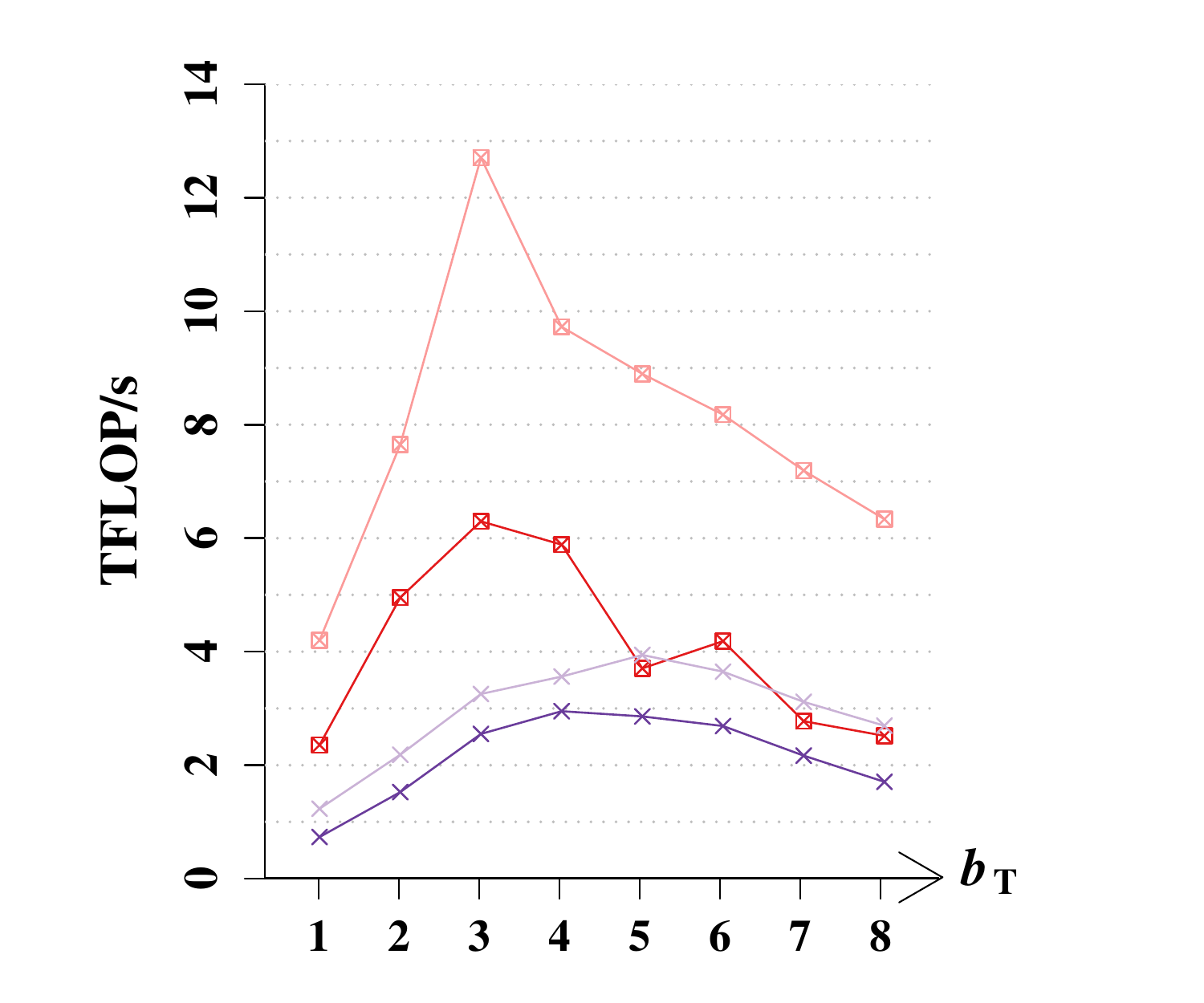}}\\
  \vspace{-0.4cm}
  \caption{Scaling with Degree of Temporal Blocking on Tesla V100 with 2D (left) and 3D (right) stencils. Float, $rad=1$.}
  \label{fig:scaling_with_tb}
  \vspace*{-0.3cm}
\end{figure}

\subsection{Scaling Performance}
Fig.~\ref{fig:scaling_with_tb} shows performance scaling with $b_{\mathrm{T}}$ on Tesla V100
when we fix the \texttt{Tuned} parameters except register limitation which is tuned for each $b_{\mathrm{T}}$.
As our model predicted, performance of 2D stencils scales up to $b_{\mathrm{T}}=10$, 3D star stencils up to $b_{\mathrm{T}}=5$,
and 3D box stencils up to $b_{\mathrm{T}}=3$. This shows that our framework is successful in optimizing different stencil types for high-degree temporal blocking,
and that our model is successful in predicting the trend of performance variations.

Fig.~\ref{fig:star-box} provides the scaling performance from first-order to fourth-order stencils on Tesla V100.
The best performance of first-order stencils is gained with high-degree temporal blocking sizes (2D: 8$\sim$15, 3D: 3$\sim$5).
For 2D stencils and 3D star stencils, most cases including fourth-order stencils achieved the best performance with $b_{\mathrm{T}}>=2$.
The only exception is high-order 3D box stencils where register pressure and the ratio of halo size to spatial block size is too high to allow performance scaling with temporal blocking.
Though for these stencils we still achieved around 60\% of the peak compute performance on Tesla V100 without temporal blocking.
It is noteworthy that our framework is the only framework that has achieved high-performance high-order stencil computation with multi-degree temporal blocking so far.
Rawat et al.~\cite{Rawat:2018:ROS:3178487.3178500} recently proposed a reordering framework to specifically target high-order stencils that benefit less from temporal blocking.
We compiled and executed the most compute-intensive 3D single-array benchmark (3d125pt) from their public repository (https://github.com/pssrawat \linebreak /ppopp-artifact)
which achieved 41\% of the peak (double-precision) compute performance on Tesla V100, while AN5D achieved 51\% of the peak without temporal blocking.
This indicates that even for high-order stencils where temporal blocking is not necessarily applicable, our framework can achieve higher computational efficiency compared to
state-of-the-art.
 \vspace*{-0.1cm}
\section{Conclusion}

In this paper we presented AN5D, our stencil framework for high-degree temporal blocking on GPUs.
With careful register-level optimizations and shared memory double-buffering,
we managed to implement temporal blocking as a practical optimization for low and mid-order stencils,
while improving computational efficiency for high-order stencils where temporal blocking is less applicable.
Moreover, our performance model allowed us to quickly choose the best configuration for each stencil pattern
and facilitate portable high performance among different GPUs.
We showed the efficiency of our framework with respect to register pressure and
performance scaling with high degrees of temporal blocking, and,
through a comprehensive comparison with previous work using typical stencil patterns,
we demonstrated that temporal blocking is crucial
to achieve high computational performance for stencil computation.

For future work, %
we plan to add support for source code transformation techniques such as warp
specialization and idle-wrap elimination to our framework to potentially enable lower
register pressure and better shared memory efficiency, and implement multi-output temporal blocking to optimize multi-statement stencils.
 \vspace*{-0.1cm}
\section*{Acknowledgement}

This work was partially supported by JST-CREST under Grant Number JPMJCR19F5. Computational resource of AI Bridging Cloud Infrastructure (ABCI) provided by National Institute of Advanced Industrial Science and Technology (AIST) was used.
 \vspace*{-0.1cm}
\balance
\bibliographystyle{ACM-Reference-Format}
\bibliography{main}


\begin{thebibliography}{41}


\ifx \showCODEN    \undefined \def \showCODEN     #1{\unskip}     \fi
\ifx \showDOI      \undefined \def \showDOI       #1{#1}\fi
\ifx \showISBNx    \undefined \def \showISBNx     #1{\unskip}     \fi
\ifx \showISBNxiii \undefined \def \showISBNxiii  #1{\unskip}     \fi
\ifx \showISSN     \undefined \def \showISSN      #1{\unskip}     \fi
\ifx \showLCCN     \undefined \def \showLCCN      #1{\unskip}     \fi
\ifx \shownote     \undefined \def \shownote      #1{#1}          \fi
\ifx \showarticletitle \undefined \def \showarticletitle #1{#1}   \fi
\ifx \showURL      \undefined \def \showURL       {\relax}        \fi
\providecommand\bibfield[2]{#2}
\providecommand\bibinfo[2]{#2}
\providecommand\natexlab[1]{#1}
\providecommand\showeprint[2][]{arXiv:#2}

\bibitem[\protect\citeauthoryear{Ao, Yang, Wang, Xue, Fu, Liu, Gan, Xu, and
  Ma}{Ao et~al\mbox{.}}{2017}]%
        {7967144}
\bibfield{author}{\bibinfo{person}{Y. Ao}, \bibinfo{person}{C. Yang},
  \bibinfo{person}{X. Wang}, \bibinfo{person}{W. Xue}, \bibinfo{person}{H. Fu},
  \bibinfo{person}{F. Liu}, \bibinfo{person}{L. Gan}, \bibinfo{person}{P. Xu},
  {and} \bibinfo{person}{W. Ma}.} \bibinfo{year}{2017}\natexlab{}.
\newblock \showarticletitle{26 PFLOPS Stencil Computations for Atmospheric
  Modeling on Sunway TaihuLight}. In \bibinfo{booktitle}{\emph{2017 IEEE
  International Parallel and Distributed Processing Symposium (IPDPS)}}.
  \bibinfo{pages}{535--544}.
\newblock
\showISSN{1530-2075}
\urldef\tempurl%
\url{https://doi.org/10.1109/IPDPS.2017.9}
\showDOI{\tempurl}


\bibitem[\protect\citeauthoryear{Bondhugula, Bandishti, and
  Pananilath}{Bondhugula et~al\mbox{.}}{2017}]%
        {bondhugula2017diamond}
\bibfield{author}{\bibinfo{person}{Uday Bondhugula}, \bibinfo{person}{Vinayaka
  Bandishti}, {and} \bibinfo{person}{Irshad Pananilath}.}
  \bibinfo{year}{2017}\natexlab{}.
\newblock \showarticletitle{Diamond Tiling: Tiling Techniques to Maximize
  Parallelism for Stencil Computations}.
\newblock \bibinfo{journal}{\emph{IEEE Trans. Parallel Distrib. Syst.}}
  \bibinfo{volume}{28}, \bibinfo{number}{5} (\bibinfo{date}{May}
  \bibinfo{year}{2017}), \bibinfo{pages}{1285--1298}.
\newblock
\showISSN{1045-9219}
\urldef\tempurl%
\url{https://doi.org/10.1109/TPDS.2016.2615094}
\showDOI{\tempurl}


\bibitem[\protect\citeauthoryear{{Bondhugula}, {Bandishti}, and
  {Pananilath}}{{Bondhugula} et~al\mbox{.}}{2017}]%
        {7582549}
\bibfield{author}{\bibinfo{person}{U. {Bondhugula}}, \bibinfo{person}{V.
  {Bandishti}}, {and} \bibinfo{person}{I. {Pananilath}}.}
  \bibinfo{year}{2017}\natexlab{}.
\newblock \showarticletitle{Diamond Tiling: Tiling Techniques to Maximize
  Parallelism for Stencil Computations}.
\newblock \bibinfo{journal}{\emph{IEEE Transactions on Parallel and Distributed
  Systems}} \bibinfo{volume}{28}, \bibinfo{number}{5} (\bibinfo{date}{May}
  \bibinfo{year}{2017}), \bibinfo{pages}{1285--1298}.
\newblock
\showISSN{1045-9219}
\urldef\tempurl%
\url{https://doi.org/10.1109/TPDS.2016.2615094}
\showDOI{\tempurl}


\bibitem[\protect\citeauthoryear{{Chi}, {Cong}, {Wei}, and {Zhou}}{{Chi}
  et~al\mbox{.}}{2018}]%
        {congstencilfpga}
\bibfield{author}{\bibinfo{person}{Y. {Chi}}, \bibinfo{person}{J. {Cong}},
  \bibinfo{person}{P. {Wei}}, {and} \bibinfo{person}{P. {Zhou}}.}
  \bibinfo{year}{2018}\natexlab{}.
\newblock \showarticletitle{SODA: Stencil with Optimized Dataflow
  Architecture}. In \bibinfo{booktitle}{\emph{2018 IEEE/ACM International
  Conference on Computer-Aided Design (ICCAD)}}. \bibinfo{pages}{1--8}.
\newblock
\showISSN{1558-2434}
\urldef\tempurl%
\url{https://doi.org/10.1145/3240765.3240850}
\showDOI{\tempurl}


\bibitem[\protect\citeauthoryear{de~Fine~Licht, Blott, and
  Hoefler}{de~Fine~Licht et~al\mbox{.}}{2018}]%
        {torstenfpgastencil}
\bibfield{author}{\bibinfo{person}{Johannes de Fine~Licht},
  \bibinfo{person}{Michaela Blott}, {and} \bibinfo{person}{Torsten Hoefler}.}
  \bibinfo{year}{2018}\natexlab{}.
\newblock \showarticletitle{Designing Scalable FPGA Architectures Using
  High-level Synthesis}. In \bibinfo{booktitle}{\emph{Proceedings of the 23rd
  ACM SIGPLAN Symposium on Principles and Practice of Parallel Programming}}
  \emph{(\bibinfo{series}{PPoPP '18})}. \bibinfo{publisher}{ACM},
  \bibinfo{address}{New York, NY, USA}, \bibinfo{pages}{403--404}.
\newblock
\showISBNx{978-1-4503-4982-6}
\urldef\tempurl%
\url{https://doi.org/10.1145/3178487.3178527}
\showDOI{\tempurl}


\bibitem[\protect\citeauthoryear{Deakin, Price, Martineau, and
  McIntosh-Smith}{Deakin et~al\mbox{.}}{2016}]%
        {babel}
\bibfield{author}{\bibinfo{person}{Tom Deakin}, \bibinfo{person}{James Price},
  \bibinfo{person}{Matt Martineau}, {and} \bibinfo{person}{Simon
  McIntosh-Smith}.} \bibinfo{year}{2016}\natexlab{}.
\newblock \showarticletitle{GPU-STREAM v2.0: Benchmarking the Achievable Memory
  Bandwidth of Many-Core Processors Across Diverse Parallel Programming
  Models}. In \bibinfo{booktitle}{\emph{High Performance Computing}},
  \bibfield{editor}{\bibinfo{person}{Michela Taufer}, \bibinfo{person}{Bernd
  Mohr}, {and} \bibinfo{person}{Julian~M. Kunkel}} (Eds.).
  \bibinfo{publisher}{Springer International Publishing},
  \bibinfo{address}{Cham}, \bibinfo{pages}{489--507}.
\newblock
\showISBNx{978-3-319-46079-6}


\bibitem[\protect\citeauthoryear{Grosser, Cohen, Holewinski, Sadayappan, and
  Verdoolaege}{Grosser et~al\mbox{.}}{2014a}]%
        {hybrid}
\bibfield{author}{\bibinfo{person}{Tobias Grosser}, \bibinfo{person}{Albert
  Cohen}, \bibinfo{person}{Justin Holewinski}, \bibinfo{person}{P. Sadayappan},
  {and} \bibinfo{person}{Sven Verdoolaege}.} \bibinfo{year}{2014}\natexlab{a}.
\newblock \showarticletitle{Hybrid Hexagonal/Classical Tiling for GPUs}. In
  \bibinfo{booktitle}{\emph{Proceedings of Annual IEEE/ACM International
  Symposium on Code Generation and Optimization}} \emph{(\bibinfo{series}{CGO
  '14})}. \bibinfo{publisher}{ACM}, \bibinfo{address}{New York, NY, USA},
  Article \bibinfo{articleno}{66}, \bibinfo{numpages}{10}~pages.
\newblock
\showISBNx{978-1-4503-2670-4}
\urldef\tempurl%
\url{https://doi.org/10.1145/2581122.2544160}
\showDOI{\tempurl}


\bibitem[\protect\citeauthoryear{Grosser, Cohen, Kelly, Ramanujam, Sadayappan,
  and Verdoolaege}{Grosser et~al\mbox{.}}{2013a}]%
        {grosser2013split}
\bibfield{author}{\bibinfo{person}{Tobias Grosser}, \bibinfo{person}{Albert
  Cohen}, \bibinfo{person}{Paul H.~J. Kelly}, \bibinfo{person}{J. Ramanujam},
  \bibinfo{person}{P. Sadayappan}, {and} \bibinfo{person}{Sven Verdoolaege}.}
  \bibinfo{year}{2013}\natexlab{a}.
\newblock \showarticletitle{Split Tiling for GPUs: Automatic Parallelization
  Using Trapezoidal Tiles}. In \bibinfo{booktitle}{\emph{Proceedings of the 6th
  Workshop on General Purpose Processor Using Graphics Processing Units}}
  \emph{(\bibinfo{series}{GPGPU-6})}. \bibinfo{publisher}{ACM},
  \bibinfo{address}{New York, NY, USA}, \bibinfo{pages}{24--31}.
\newblock
\showISBNx{978-1-4503-2017-7}
\urldef\tempurl%
\url{https://doi.org/10.1145/2458523.2458526}
\showDOI{\tempurl}


\bibitem[\protect\citeauthoryear{Grosser, Verdoolaege, Cohen, and
  Sadayappan}{Grosser et~al\mbox{.}}{2013b}]%
        {grosser2013promises}
\bibfield{author}{\bibinfo{person}{Tobias Grosser}, \bibinfo{person}{Sven
  Verdoolaege}, \bibinfo{person}{Albert Cohen}, {and} \bibinfo{person}{P.
  Sadayappan}.} \bibinfo{year}{2013}\natexlab{b}.
\newblock \bibinfo{booktitle}{\emph{{The Promises of Hybrid Hexagonal/Classical
  Tiling for GPU}}}.
\newblock \bibinfo{type}{Research Report} RR-8339.
  \bibinfo{institution}{{INRIA}}.
\newblock
\urldef\tempurl%
\url{https://hal.inria.fr/hal-00848691}
\showURL{%
\tempurl}


\bibitem[\protect\citeauthoryear{Grosser, Verdoolaege, Cohen, and
  Sadayappan}{Grosser et~al\mbox{.}}{2014b}]%
        {grosser2014relation}
\bibfield{author}{\bibinfo{person}{Tobias Grosser}, \bibinfo{person}{Sven
  Verdoolaege}, \bibinfo{person}{Albert Cohen}, {and} \bibinfo{person}{P.
  Sadayappan}.} \bibinfo{year}{2014}\natexlab{b}.
\newblock \showarticletitle{The Relation Between Diamond Tiling and Hexagonal
  Tiling}.
\newblock \bibinfo{journal}{\emph{Parallel Processing Letters}}
  \bibinfo{volume}{24}, \bibinfo{number}{03} (\bibinfo{year}{2014}),
  \bibinfo{pages}{1441002}.
\newblock
\urldef\tempurl%
\url{https://doi.org/10.1142/S0129626414410023}
\showDOI{\tempurl}
\showeprint{https://doi.org/10.1142/S0129626414410023}


\bibitem[\protect\citeauthoryear{Hagedorn, Stoltzfus, Steuwer, Gorlatch, and
  Dubach}{Hagedorn et~al\mbox{.}}{2018}]%
        {Hagedorn:2018:HPS:3179541.3168824}
\bibfield{author}{\bibinfo{person}{Bastian Hagedorn}, \bibinfo{person}{Larisa
  Stoltzfus}, \bibinfo{person}{Michel Steuwer}, \bibinfo{person}{Sergei
  Gorlatch}, {and} \bibinfo{person}{Christophe Dubach}.}
  \bibinfo{year}{2018}\natexlab{}.
\newblock \showarticletitle{High Performance Stencil Code Generation with
  Lift}. In \bibinfo{booktitle}{\emph{Proceedings of the 2018 International
  Symposium on Code Generation and Optimization}} \emph{(\bibinfo{series}{CGO
  2018})}. \bibinfo{publisher}{ACM}, \bibinfo{address}{New York, NY, USA},
  \bibinfo{pages}{100--112}.
\newblock
\showISBNx{978-1-4503-5617-6}
\urldef\tempurl%
\url{https://doi.org/10.1145/3168824}
\showDOI{\tempurl}


\bibitem[\protect\citeauthoryear{Holewinski, Pouchet, and
  Sadayappan}{Holewinski et~al\mbox{.}}{2012}]%
        {Holewinski:2012:HCG:2304576.2304619}
\bibfield{author}{\bibinfo{person}{Justin Holewinski},
  \bibinfo{person}{Louis-No\"{e}l Pouchet}, {and} \bibinfo{person}{P.
  Sadayappan}.} \bibinfo{year}{2012}\natexlab{}.
\newblock \showarticletitle{High-performance Code Generation for Stencil
  Computations on GPU Architectures}. In \bibinfo{booktitle}{\emph{Proceedings
  of the 26th ACM International Conference on Supercomputing}}
  \emph{(\bibinfo{series}{ICS '12})}. \bibinfo{publisher}{ACM},
  \bibinfo{address}{New York, NY, USA}, \bibinfo{pages}{311--320}.
\newblock
\showISBNx{978-1-4503-1316-2}
\urldef\tempurl%
\url{https://doi.org/10.1145/2304576.2304619}
\showDOI{\tempurl}


\bibitem[\protect\citeauthoryear{Irigoin and Triolet}{Irigoin and
  Triolet}{1988}]%
        {irigoin1988supernode}
\bibfield{author}{\bibinfo{person}{F. Irigoin} {and} \bibinfo{person}{R.
  Triolet}.} \bibinfo{year}{1988}\natexlab{}.
\newblock \showarticletitle{Supernode Partitioning}. In
  \bibinfo{booktitle}{\emph{Proceedings of the 15th ACM SIGPLAN-SIGACT
  Symposium on Principles of Programming Languages}}
  \emph{(\bibinfo{series}{POPL '88})}. \bibinfo{publisher}{ACM},
  \bibinfo{address}{New York, NY, USA}, \bibinfo{pages}{319--329}.
\newblock
\showISBNx{0-89791-252-7}
\urldef\tempurl%
\url{https://doi.org/10.1145/73560.73588}
\showDOI{\tempurl}


\bibitem[\protect\citeauthoryear{Kamil, Datta, Williams, Oliker, Shalf, and
  Yelick}{Kamil et~al\mbox{.}}{2006}]%
        {Kamil:2006:IEO:1178597.1178605}
\bibfield{author}{\bibinfo{person}{Shoaib Kamil}, \bibinfo{person}{Kaushik
  Datta}, \bibinfo{person}{Samuel Williams}, \bibinfo{person}{Leonid Oliker},
  \bibinfo{person}{John Shalf}, {and} \bibinfo{person}{Katherine Yelick}.}
  \bibinfo{year}{2006}\natexlab{}.
\newblock \showarticletitle{Implicit and Explicit Optimizations for Stencil
  Computations}. In \bibinfo{booktitle}{\emph{Proceedings of the 2006 Workshop
  on Memory System Performance and Correctness}} \emph{(\bibinfo{series}{MSPC
  '06})}. \bibinfo{publisher}{ACM}, \bibinfo{address}{New York, NY, USA},
  \bibinfo{pages}{51--60}.
\newblock
\showISBNx{1-59593-578-9}
\urldef\tempurl%
\url{https://doi.org/10.1145/1178597.1178605}
\showDOI{\tempurl}


\bibitem[\protect\citeauthoryear{{Konstantinidis} and
  {Cotronis}}{{Konstantinidis} and {Cotronis}}{2016}]%
        {gpumembench}
\bibfield{author}{\bibinfo{person}{E. {Konstantinidis}} {and}
  \bibinfo{person}{Y. {Cotronis}}.} \bibinfo{year}{2016}\natexlab{}.
\newblock \showarticletitle{A Quantitative Performance Evaluation of Fast
  on-Chip Memories of GPUs}. In \bibinfo{booktitle}{\emph{2016 24th Euromicro
  International Conference on Parallel, Distributed, and Network-Based
  Processing (PDP)}}. \bibinfo{pages}{448--455}.
\newblock
\showISSN{2377-5750}
\urldef\tempurl%
\url{https://doi.org/10.1109/PDP.2016.56}
\showDOI{\tempurl}


\bibitem[\protect\citeauthoryear{Krishnamoorthy, Baskaran, Bondhugula,
  Ramanujam, Rountev, and Sadayappan}{Krishnamoorthy et~al\mbox{.}}{2007}]%
        {Krishnamoorthy:2007:EAP:1250734.1250761}
\bibfield{author}{\bibinfo{person}{Sriram Krishnamoorthy},
  \bibinfo{person}{Muthu Baskaran}, \bibinfo{person}{Uday Bondhugula},
  \bibinfo{person}{J. Ramanujam}, \bibinfo{person}{Atanas Rountev}, {and}
  \bibinfo{person}{P Sadayappan}.} \bibinfo{year}{2007}\natexlab{}.
\newblock \showarticletitle{Effective Automatic Parallelization of Stencil
  Computations}. In \bibinfo{booktitle}{\emph{Proceedings of the 28th ACM
  SIGPLAN Conference on Programming Language Design and Implementation}}
  \emph{(\bibinfo{series}{PLDI '07})}. \bibinfo{publisher}{ACM},
  \bibinfo{address}{New York, NY, USA}, \bibinfo{pages}{235--244}.
\newblock
\showISBNx{978-1-59593-633-2}
\urldef\tempurl%
\url{https://doi.org/10.1145/1250734.1250761}
\showDOI{\tempurl}


\bibitem[\protect\citeauthoryear{Maruyama and Aoki}{Maruyama and Aoki}{2014}]%
        {maruyama2014optimizing}
\bibfield{author}{\bibinfo{person}{Naoya Maruyama} {and}
  \bibinfo{person}{Takayuki Aoki}.} \bibinfo{year}{2014}\natexlab{}.
\newblock \showarticletitle{{O}ptimizing {S}tencil {C}omputations for {NVIDIA}
  {K}epler {GPU}s}. In \bibinfo{booktitle}{\emph{{P}roceedings of the 1st
  {I}nternational {W}orkshop on {H}igh-{P}erformance {S}tencil
  {C}omputations}}, \bibfield{editor}{\bibinfo{person}{Armin
  Gr{\"o}{\ss}linger} {and} \bibinfo{person}{Harald K{\"o}stler}} (Eds.).
  \bibinfo{address}{Vienna, Austria}, \bibinfo{pages}{89--95}.
\newblock
\urldef\tempurl%
\url{http://www.exastencils.org/histencils/2014/}
\showURL{%
\tempurl}


\bibitem[\protect\citeauthoryear{Meng and Skadron}{Meng and Skadron}{2009}]%
        {Meng:2009:PMA:1542275.1542313}
\bibfield{author}{\bibinfo{person}{Jiayuan Meng} {and} \bibinfo{person}{Kevin
  Skadron}.} \bibinfo{year}{2009}\natexlab{}.
\newblock \showarticletitle{Performance Modeling and Automatic Ghost Zone
  Optimization for Iterative Stencil Loops on GPUs}. In
  \bibinfo{booktitle}{\emph{Proceedings of the 23rd International Conference on
  Supercomputing}} \emph{(\bibinfo{series}{ICS '09})}.
  \bibinfo{publisher}{ACM}, \bibinfo{address}{New York, NY, USA},
  \bibinfo{pages}{256--265}.
\newblock
\showISBNx{978-1-60558-498-0}
\urldef\tempurl%
\url{https://doi.org/10.1145/1542275.1542313}
\showDOI{\tempurl}


\bibitem[\protect\citeauthoryear{Muranushi and Makino}{Muranushi and
  Makino}{2015}]%
        {muranushi2015optimal}
\bibfield{author}{\bibinfo{person}{Takayuki Muranushi} {and}
  \bibinfo{person}{Junichiro Makino}.} \bibinfo{year}{2015}\natexlab{}.
\newblock \showarticletitle{Optimal Temporal Blocking for Stencil Computation}.
\newblock \bibinfo{journal}{\emph{Procedia Computer Science}}
  \bibinfo{volume}{51} (\bibinfo{year}{2015}), \bibinfo{pages}{1303 -- 1312}.
\newblock
\showISSN{1877-0509}
\urldef\tempurl%
\url{https://doi.org/10.1016/j.procs.2015.05.315}
\showDOI{\tempurl}
\newblock
\shownote{International Conference On Computational Science, ICCS 2015.}


\bibitem[\protect\citeauthoryear{Nguyen, Satish, Chhugani, Kim, and
  Dubey}{Nguyen et~al\mbox{.}}{2010}]%
        {3.5d}
\bibfield{author}{\bibinfo{person}{A. Nguyen}, \bibinfo{person}{N. Satish},
  \bibinfo{person}{J. Chhugani}, \bibinfo{person}{C. Kim}, {and}
  \bibinfo{person}{P. Dubey}.} \bibinfo{year}{2010}\natexlab{}.
\newblock \showarticletitle{3.5-D Blocking Optimization for Stencil
  Computations on Modern CPUs and GPUs}. In \bibinfo{booktitle}{\emph{SC '10:
  Proceedings of the 2010 ACM/IEEE International Conference for High
  Performance Computing, Networking, Storage and Analysis}}.
  \bibinfo{pages}{1--13}.
\newblock
\showISSN{2167-4329}
\urldef\tempurl%
\url{https://doi.org/10.1109/SC.2010.2}
\showDOI{\tempurl}


\bibitem[\protect\citeauthoryear{Prajapati, Ranasinghe, Rajopadhye, Andonov,
  Djidjev, and Grosser}{Prajapati et~al\mbox{.}}{2017}]%
        {prajapati2017simple}
\bibfield{author}{\bibinfo{person}{Nirmal Prajapati}, \bibinfo{person}{Waruna
  Ranasinghe}, \bibinfo{person}{Sanjay Rajopadhye}, \bibinfo{person}{Rumen
  Andonov}, \bibinfo{person}{Hristo Djidjev}, {and} \bibinfo{person}{Tobias
  Grosser}.} \bibinfo{year}{2017}\natexlab{}.
\newblock \showarticletitle{Simple, Accurate, Analytical Time Modeling and
  Optimal Tile Size Selection for GPGPU Stencils}. In
  \bibinfo{booktitle}{\emph{Proceedings of the 22Nd ACM SIGPLAN Symposium on
  Principles and Practice of Parallel Programming}}
  \emph{(\bibinfo{series}{PPoPP '17})}. \bibinfo{publisher}{ACM},
  \bibinfo{address}{New York, NY, USA}, \bibinfo{pages}{163--177}.
\newblock
\showISBNx{978-1-4503-4493-7}
\urldef\tempurl%
\url{https://doi.org/10.1145/3018743.3018744}
\showDOI{\tempurl}


\bibitem[\protect\citeauthoryear{Ravishankar, Holewinski, and
  Grover}{Ravishankar et~al\mbox{.}}{2015}]%
        {Ravishankar:2015:FDI:2716282.2716290}
\bibfield{author}{\bibinfo{person}{Mahesh Ravishankar}, \bibinfo{person}{Justin
  Holewinski}, {and} \bibinfo{person}{Vinod Grover}.}
  \bibinfo{year}{2015}\natexlab{}.
\newblock \showarticletitle{Forma: A DSL for Image Processing Applications to
  Target GPUs and Multi-core CPUs}. In \bibinfo{booktitle}{\emph{Proceedings of
  the 8th Workshop on General Purpose Processing Using GPUs}}
  \emph{(\bibinfo{series}{GPGPU-8})}. \bibinfo{publisher}{ACM},
  \bibinfo{address}{New York, NY, USA}, \bibinfo{pages}{109--120}.
\newblock
\showISBNx{978-1-4503-3407-5}
\urldef\tempurl%
\url{https://doi.org/10.1145/2716282.2716290}
\showDOI{\tempurl}


\bibitem[\protect\citeauthoryear{Rawat, Kong, Henretty, Holewinski, Stock,
  Pouchet, Ramanujam, Rountev, and Sadayappan}{Rawat et~al\mbox{.}}{2015}]%
        {Rawat:2015:SMD:2830018.2830025}
\bibfield{author}{\bibinfo{person}{Prashant Rawat}, \bibinfo{person}{Martin
  Kong}, \bibinfo{person}{Tom Henretty}, \bibinfo{person}{Justin Holewinski},
  \bibinfo{person}{Kevin Stock}, \bibinfo{person}{Louis-No\"{e}l Pouchet},
  \bibinfo{person}{J. Ramanujam}, \bibinfo{person}{Atanas Rountev}, {and}
  \bibinfo{person}{P. Sadayappan}.} \bibinfo{year}{2015}\natexlab{}.
\newblock \showarticletitle{SDSLc: A Multi-target Domain-specific Compiler for
  Stencil Computations}. In \bibinfo{booktitle}{\emph{Proceedings of the 5th
  International Workshop on Domain-Specific Languages and High-Level Frameworks
  for High Performance Computing}} \emph{(\bibinfo{series}{WOLFHPC '15})}.
  \bibinfo{publisher}{ACM}, \bibinfo{address}{New York, NY, USA}, Article
  \bibinfo{articleno}{6}, \bibinfo{numpages}{10}~pages.
\newblock
\showISBNx{978-1-4503-4016-8}
\urldef\tempurl%
\url{https://doi.org/10.1145/2830018.2830025}
\showDOI{\tempurl}


\bibitem[\protect\citeauthoryear{Rawat, Hong, Ravishankar, Grover, Pouchet, and
  Sadayappan}{Rawat et~al\mbox{.}}{2016}]%
        {stencilgen1}
\bibfield{author}{\bibinfo{person}{Prashant~Singh Rawat},
  \bibinfo{person}{Changwan Hong}, \bibinfo{person}{Mahesh Ravishankar},
  \bibinfo{person}{Vinod Grover}, \bibinfo{person}{Louis-No\"{e}l Pouchet},
  {and} \bibinfo{person}{P. Sadayappan}.} \bibinfo{year}{2016}\natexlab{}.
\newblock \showarticletitle{Effective Resource Management for Enhancing
  Performance of 2D and 3D Stencils on GPUs}. In
  \bibinfo{booktitle}{\emph{Proceedings of the 9th Annual Workshop on General
  Purpose Processing Using Graphics Processing Unit}}
  \emph{(\bibinfo{series}{GPGPU '16})}. \bibinfo{publisher}{ACM},
  \bibinfo{address}{New York, NY, USA}, \bibinfo{pages}{92--102}.
\newblock
\showISBNx{978-1-4503-4195-0}
\urldef\tempurl%
\url{https://doi.org/10.1145/2884045.2884047}
\showDOI{\tempurl}


\bibitem[\protect\citeauthoryear{Rawat, Rastello, Sukumaran-Rajam, Pouchet,
  Rountev, and Sadayappan}{Rawat et~al\mbox{.}}{2018a}]%
        {Rawat:2018:ROS:3178487.3178500}
\bibfield{author}{\bibinfo{person}{Prashant~Singh Rawat},
  \bibinfo{person}{Fabrice Rastello}, \bibinfo{person}{Aravind
  Sukumaran-Rajam}, \bibinfo{person}{Louis-No\"{e}l Pouchet},
  \bibinfo{person}{Atanas Rountev}, {and} \bibinfo{person}{P. Sadayappan}.}
  \bibinfo{year}{2018}\natexlab{a}.
\newblock \showarticletitle{Register Optimizations for Stencils on GPUs}. In
  \bibinfo{booktitle}{\emph{Proceedings of the 23rd ACM SIGPLAN Symposium on
  Principles and Practice of Parallel Programming}}
  \emph{(\bibinfo{series}{PPoPP '18})}. \bibinfo{publisher}{ACM},
  \bibinfo{address}{New York, NY, USA}, \bibinfo{pages}{168--182}.
\newblock
\showISBNx{978-1-4503-4982-6}
\urldef\tempurl%
\url{https://doi.org/10.1145/3178487.3178500}
\showDOI{\tempurl}


\bibitem[\protect\citeauthoryear{Rawat, Vaidya, Sukumaran-Rajam, Ravishankar,
  Grover, Rountev, Pouchet, and Sadayappan}{Rawat et~al\mbox{.}}{2018b}]%
        {stencilgen2}
\bibfield{author}{\bibinfo{person}{P.~S. Rawat}, \bibinfo{person}{M. Vaidya},
  \bibinfo{person}{A. Sukumaran-Rajam}, \bibinfo{person}{M. Ravishankar},
  \bibinfo{person}{V. Grover}, \bibinfo{person}{A. Rountev},
  \bibinfo{person}{L. Pouchet}, {and} \bibinfo{person}{P. Sadayappan}.}
  \bibinfo{year}{2018}\natexlab{b}.
\newblock \showarticletitle{Domain-Specific Optimization and Generation of
  High-Performance GPU Code for Stencil Computations}.
\newblock \bibinfo{journal}{\emph{Proc. IEEE}} \bibinfo{volume}{106},
  \bibinfo{number}{11} (\bibinfo{date}{Nov} \bibinfo{year}{2018}),
  \bibinfo{pages}{1902--1920}.
\newblock
\showISSN{0018-9219}
\urldef\tempurl%
\url{https://doi.org/10.1109/JPROC.2018.2862896}
\showDOI{\tempurl}


\bibitem[\protect\citeauthoryear{Rawat, Vaidya, Sukumaran-Rajam, Rountev,
  Pouchet, and Sadayappan}{Rawat et~al\mbox{.}}{2019}]%
        {rawatoptimizing}
\bibfield{author}{\bibinfo{person}{Prashant~Singh Rawat},
  \bibinfo{person}{Miheer Vaidya}, \bibinfo{person}{Aravind Sukumaran-Rajam},
  \bibinfo{person}{Atanas Rountev}, \bibinfo{person}{Louis-No{\"e}l Pouchet},
  {and} \bibinfo{person}{P Sadayappan}.} \bibinfo{year}{2019}\natexlab{}.
\newblock \showarticletitle{On Optimizing Complex Stencils on GPUs}. In
  \bibinfo{booktitle}{\emph{2019 IEEE International Parallel and Distributed
  Processing Symposium (IPDPS)}}.
\newblock


\bibitem[\protect\citeauthoryear{Rossinelli, Hejazialhosseini, Hadjidoukas,
  Bekas, Curioni, Bertsch, Futral, Schmidt, Adams, and Koumoutsakos}{Rossinelli
  et~al\mbox{.}}{2013}]%
        {Rossinelli:2013:PSC:2503210.2504565}
\bibfield{author}{\bibinfo{person}{Diego Rossinelli}, \bibinfo{person}{Babak
  Hejazialhosseini}, \bibinfo{person}{Panagiotis Hadjidoukas},
  \bibinfo{person}{Costas Bekas}, \bibinfo{person}{Alessandro Curioni},
  \bibinfo{person}{Adam Bertsch}, \bibinfo{person}{Scott Futral},
  \bibinfo{person}{Steffen~J. Schmidt}, \bibinfo{person}{Nikolaus~A. Adams},
  {and} \bibinfo{person}{Petros Koumoutsakos}.}
  \bibinfo{year}{2013}\natexlab{}.
\newblock \showarticletitle{11 PFLOP/s Simulations of Cloud Cavitation
  Collapse}. In \bibinfo{booktitle}{\emph{Proceedings of the International
  Conference on High Performance Computing, Networking, Storage and Analysis}}
  \emph{(\bibinfo{series}{SC '13})}. \bibinfo{publisher}{ACM},
  \bibinfo{address}{New York, NY, USA}, Article \bibinfo{articleno}{3},
  \bibinfo{numpages}{13}~pages.
\newblock
\showISBNx{978-1-4503-2378-9}
\urldef\tempurl%
\url{https://doi.org/10.1145/2503210.2504565}
\showDOI{\tempurl}


\bibitem[\protect\citeauthoryear{Shimokawabe, Aoki, Muroi, Ishida, Kawano,
  Endo, Nukada, Maruyama, and Matsuoka}{Shimokawabe et~al\mbox{.}}{2010}]%
        {shimokawabe201080}
\bibfield{author}{\bibinfo{person}{Takashi Shimokawabe},
  \bibinfo{person}{Takayuki Aoki}, \bibinfo{person}{Chiashi Muroi},
  \bibinfo{person}{Junichi Ishida}, \bibinfo{person}{Kohei Kawano},
  \bibinfo{person}{Toshio Endo}, \bibinfo{person}{Akira Nukada},
  \bibinfo{person}{Naoya Maruyama}, {and} \bibinfo{person}{Satoshi Matsuoka}.}
  \bibinfo{year}{2010}\natexlab{}.
\newblock \showarticletitle{An 80-Fold Speedup, 15.0 TFlops Full GPU
  Acceleration of Non-Hydrostatic Weather Model ASUCA Production Code}. In
  \bibinfo{booktitle}{\emph{Proceedings of the 2010 ACM/IEEE International
  Conference for High Performance Computing, Networking, Storage and Analysis}}
  \emph{(\bibinfo{series}{SC '10})}. \bibinfo{publisher}{IEEE Computer
  Society}, \bibinfo{address}{Washington, DC, USA}, \bibinfo{pages}{1--11}.
\newblock
\showISBNx{978-1-4244-7559-9}
\urldef\tempurl%
\url{https://doi.org/10.1109/SC.2010.9}
\showDOI{\tempurl}


\bibitem[\protect\citeauthoryear{Shimokawabe, Aoki, Takaki, Endo, Yamanaka,
  Maruyama, Nukada, and Matsuoka}{Shimokawabe et~al\mbox{.}}{2011}]%
        {shimokawabe2011peta}
\bibfield{author}{\bibinfo{person}{Takashi Shimokawabe},
  \bibinfo{person}{Takayuki Aoki}, \bibinfo{person}{Tomohiro Takaki},
  \bibinfo{person}{Toshio Endo}, \bibinfo{person}{Akinori Yamanaka},
  \bibinfo{person}{Naoya Maruyama}, \bibinfo{person}{Akira Nukada}, {and}
  \bibinfo{person}{Satoshi Matsuoka}.} \bibinfo{year}{2011}\natexlab{}.
\newblock \showarticletitle{Peta-scale Phase-field Simulation for Dendritic
  Solidification on the TSUBAME 2.0 Supercomputer}. In
  \bibinfo{booktitle}{\emph{Proceedings of 2011 International Conference for
  High Performance Computing, Networking, Storage and Analysis}}
  \emph{(\bibinfo{series}{SC '11})}. \bibinfo{publisher}{ACM},
  \bibinfo{address}{New York, NY, USA}, Article \bibinfo{articleno}{3},
  \bibinfo{numpages}{11}~pages.
\newblock
\showISBNx{978-1-4503-0771-0}
\urldef\tempurl%
\url{https://doi.org/10.1145/2063384.2063388}
\showDOI{\tempurl}


\bibitem[\protect\citeauthoryear{Tang, Tan, Krishnamoorthy, Wong, Kuo, Goh,
  Turner, and Wong}{Tang et~al\mbox{.}}{2013}]%
        {6569833}
\bibfield{author}{\bibinfo{person}{W.~T. Tang}, \bibinfo{person}{W.~J. Tan},
  \bibinfo{person}{R. Krishnamoorthy}, \bibinfo{person}{Y.~W. Wong},
  \bibinfo{person}{S. Kuo}, \bibinfo{person}{R.~S.~M. Goh},
  \bibinfo{person}{S.~J. Turner}, {and} \bibinfo{person}{W. Wong}.}
  \bibinfo{year}{2013}\natexlab{}.
\newblock \showarticletitle{Optimizing and Auto-Tuning Iterative Stencil Loops
  for GPUs with the In-Plane Method}. In \bibinfo{booktitle}{\emph{2013 IEEE
  27th International Symposium on Parallel and Distributed Processing}}.
  \bibinfo{pages}{452--462}.
\newblock
\showISSN{1530-2075}
\urldef\tempurl%
\url{https://doi.org/10.1109/IPDPS.2013.79}
\showDOI{\tempurl}


\bibitem[\protect\citeauthoryear{Tang, Chowdhury, Kuszmaul, Luk, and
  Leiserson}{Tang et~al\mbox{.}}{2011}]%
        {Tang:2011:PSC:1989493.1989508}
\bibfield{author}{\bibinfo{person}{Yuan Tang}, \bibinfo{person}{Rezaul~Alam
  Chowdhury}, \bibinfo{person}{Bradley~C. Kuszmaul}, \bibinfo{person}{Chi-Keung
  Luk}, {and} \bibinfo{person}{Charles~E. Leiserson}.}
  \bibinfo{year}{2011}\natexlab{}.
\newblock \showarticletitle{The Pochoir Stencil Compiler}. In
  \bibinfo{booktitle}{\emph{Proceedings of the Twenty-third Annual ACM
  Symposium on Parallelism in Algorithms and Architectures}}
  \emph{(\bibinfo{series}{SPAA '11})}. \bibinfo{publisher}{ACM},
  \bibinfo{address}{New York, NY, USA}, \bibinfo{pages}{117--128}.
\newblock
\showISBNx{978-1-4503-0743-7}
\urldef\tempurl%
\url{https://doi.org/10.1145/1989493.1989508}
\showDOI{\tempurl}


\bibitem[\protect\citeauthoryear{TOP500.org}{TOP500.org}{2018}]%
        {top500}
\bibfield{author}{\bibinfo{person}{TOP500.org}.}
  \bibinfo{year}{2018}\natexlab{}.
\newblock \bibinfo{title}{November 2018 | TOP500 Supercomputer Sites}.
\newblock
\newblock
\urldef\tempurl%
\url{https://www.top500.org/lists/2018/11/}
\showURL{%
\tempurl}


\bibitem[\protect\citeauthoryear{Verdoolaege}{Verdoolaege}{2010}]%
        {verdoolaege2010isl}
\bibfield{author}{\bibinfo{person}{Sven Verdoolaege}.}
  \bibinfo{year}{2010}\natexlab{}.
\newblock \showarticletitle{isl: An Integer Set Library for the Polyhedral
  Model}. In \bibinfo{booktitle}{\emph{Mathematical Software -- ICMS 2010}},
  \bibfield{editor}{\bibinfo{person}{Komei Fukuda}, \bibinfo{person}{Joris
  van~der Hoeven}, \bibinfo{person}{Michael Joswig}, {and}
  \bibinfo{person}{Nobuki Takayama}} (Eds.). \bibinfo{publisher}{Springer
  Berlin Heidelberg}, \bibinfo{address}{Berlin, Heidelberg},
  \bibinfo{pages}{299--302}.
\newblock
\urldef\tempurl%
\url{https://doi.org/10.1007/978-3-642-15582-6_49}
\showDOI{\tempurl}


\bibitem[\protect\citeauthoryear{Verdoolaege, Carlos~Juega, Cohen,
  Ignacio~G\'{o}mez, Tenllado, and Catthoor}{Verdoolaege et~al\mbox{.}}{2013}]%
        {verdoolaege2013polyhedral}
\bibfield{author}{\bibinfo{person}{Sven Verdoolaege}, \bibinfo{person}{Juan
  Carlos~Juega}, \bibinfo{person}{Albert Cohen}, \bibinfo{person}{Jos{\'e}
  Ignacio~G\'{o}mez}, \bibinfo{person}{Christian Tenllado}, {and}
  \bibinfo{person}{Francky Catthoor}.} \bibinfo{year}{2013}\natexlab{}.
\newblock \showarticletitle{Polyhedral Parallel Code Generation for CUDA}.
\newblock \bibinfo{journal}{\emph{ACM Trans. Archit. Code Optim.}}
  \bibinfo{volume}{9}, \bibinfo{number}{4}, Article \bibinfo{articleno}{54}
  (\bibinfo{date}{Jan.} \bibinfo{year}{2013}), \bibinfo{numpages}{23}~pages.
\newblock
\showISSN{1544-3566}
\urldef\tempurl%
\url{https://doi.org/10.1145/2400682.2400713}
\showDOI{\tempurl}


\bibitem[\protect\citeauthoryear{Verdoolaege and Grosser}{Verdoolaege and
  Grosser}{2012}]%
        {verdoolaege2012polyhedral}
\bibfield{author}{\bibinfo{person}{Sven Verdoolaege} {and}
  \bibinfo{person}{Tobias Grosser}.} \bibinfo{year}{2012}\natexlab{}.
\newblock \showarticletitle{Polyhedral Extraction Tool}. In
  \bibinfo{booktitle}{\emph{Second International Workshop on Polyhedral
  Compilation Techniques (IMPACT'12)}}. \bibinfo{address}{Paris, France}.
\newblock
\urldef\tempurl%
\url{http://impact.gforge.inria.fr/impact2012/}
\showURL{%
\tempurl}


\bibitem[\protect\citeauthoryear{Williams, Waterman, and Patterson}{Williams
  et~al\mbox{.}}{2009}]%
        {roofline}
\bibfield{author}{\bibinfo{person}{Samuel Williams}, \bibinfo{person}{Andrew
  Waterman}, {and} \bibinfo{person}{David Patterson}.}
  \bibinfo{year}{2009}\natexlab{}.
\newblock \showarticletitle{Roofline: An Insightful Visual Performance Model
  for Multicore Architectures}.
\newblock \bibinfo{journal}{\emph{Commun. ACM}} \bibinfo{volume}{52},
  \bibinfo{number}{4} (\bibinfo{date}{April} \bibinfo{year}{2009}),
  \bibinfo{pages}{65--76}.
\newblock
\showISSN{0001-0782}
\urldef\tempurl%
\url{https://doi.org/10.1145/1498765.1498785}
\showDOI{\tempurl}


\bibitem[\protect\citeauthoryear{Wolfe}{Wolfe}{1989}]%
        {Wolfe:1989:MIS:76263.76337}
\bibfield{author}{\bibinfo{person}{M. Wolfe}.} \bibinfo{year}{1989}\natexlab{}.
\newblock \showarticletitle{More Iteration Space Tiling}. In
  \bibinfo{booktitle}{\emph{Proceedings of the 1989 ACM/IEEE Conference on
  Supercomputing}} \emph{(\bibinfo{series}{Supercomputing '89})}.
  \bibinfo{publisher}{ACM}, \bibinfo{address}{New York, NY, USA},
  \bibinfo{pages}{655--664}.
\newblock
\showISBNx{0-89791-341-8}
\urldef\tempurl%
\url{https://doi.org/10.1145/76263.76337}
\showDOI{\tempurl}


\bibitem[\protect\citeauthoryear{{Yount}, {Tobin}, {Breuer}, and
  {Duran}}{{Yount} et~al\mbox{.}}{2016}]%
        {7836083}
\bibfield{author}{\bibinfo{person}{C. {Yount}}, \bibinfo{person}{J. {Tobin}},
  \bibinfo{person}{A. {Breuer}}, {and} \bibinfo{person}{A. {Duran}}.}
  \bibinfo{year}{2016}\natexlab{}.
\newblock \showarticletitle{YASK—Yet Another Stencil Kernel: A Framework for
  HPC Stencil Code-Generation and Tuning}. In \bibinfo{booktitle}{\emph{2016
  Sixth International Workshop on Domain-Specific Languages and High-Level
  Frameworks for High Performance Computing (WOLFHPC)}}.
  \bibinfo{pages}{30--39}.
\newblock
\urldef\tempurl%
\url{https://doi.org/10.1109/WOLFHPC.2016.08}
\showDOI{\tempurl}


\bibitem[\protect\citeauthoryear{Zohouri, Podobas, and Matsuoka}{Zohouri
  et~al\mbox{.}}{2018a}]%
        {zohouri2018combined}
\bibfield{author}{\bibinfo{person}{Hamid~Reza Zohouri}, \bibinfo{person}{Artur
  Podobas}, {and} \bibinfo{person}{Satoshi Matsuoka}.}
  \bibinfo{year}{2018}\natexlab{a}.
\newblock \showarticletitle{Combined Spatial and Temporal Blocking for
  High-Performance Stencil Computation on FPGAs Using OpenCL}. In
  \bibinfo{booktitle}{\emph{Proceedings of the 2018 ACM/SIGDA International
  Symposium on Field-Programmable Gate Arrays}} \emph{(\bibinfo{series}{FPGA
  '18})}. \bibinfo{publisher}{ACM}, \bibinfo{address}{New York, NY, USA},
  \bibinfo{pages}{153--162}.
\newblock
\showISBNx{978-1-4503-5614-5}
\urldef\tempurl%
\url{https://doi.org/10.1145/3174243.3174248}
\showDOI{\tempurl}


\bibitem[\protect\citeauthoryear{Zohouri, Podobas, and Matsuoka}{Zohouri
  et~al\mbox{.}}{2018b}]%
        {zohouri2018high}
\bibfield{author}{\bibinfo{person}{H.~R. Zohouri}, \bibinfo{person}{A.
  Podobas}, {and} \bibinfo{person}{S. Matsuoka}.}
  \bibinfo{year}{2018}\natexlab{b}.
\newblock \showarticletitle{High-Performance High-Order Stencil Computation on
  FPGAs Using OpenCL}. In \bibinfo{booktitle}{\emph{2018 IEEE International
  Parallel and Distributed Processing Symposium Workshops (IPDPSW)}}.
  \bibinfo{pages}{123--130}.
\newblock
\urldef\tempurl%
\url{https://doi.org/10.1109/IPDPSW.2018.00027}
\showDOI{\tempurl}


\end{thebibliography}

\end{document}